\documentclass[journal=jacsat,manuscript=article]{achemso}

\usepackage[version=3]{mhchem} 
\usepackage{natbib}
\usepackage{graphicx}
\usepackage{amsmath}
\usepackage{multirow,makecell}
\usepackage{bm}
\usepackage{float}
\usepackage{soul}
\usepackage{xcolor}
\usepackage{subcaption}
\usepackage{mciteplus}
\usepackage{siunitx}
\usepackage{adjustbox}
\usepackage{threeparttable}
\usepackage{caption}
\usepackage{indentfirst}
\usepackage{textcomp}

\usepackage{xr-hyper}
\usepackage[colorlinks,allcolors=blue]{hyperref}
\usepackage[nameinlink,noabbrev,capitalize]{cleveref}

\SectionNumbersOn

\author{Xiaoyu Wang}
\affiliation{Chemical Sciences and Engineering Division, Argonne National Laboratory, 9700 S Cass Ave, Lemont, IL 60439}
\email{xiaoyu.wang@anl.gov}
\author{Michael J. Servis}
\affiliation{Chemical Sciences and Engineering Division, Argonne National Laboratory, 9700 S Cass Ave, Lemont, IL 60439}
\email{mservis@anl.gov}

\title{Using Metadynamics to Reveal Extractant Conformational Free Energy Landscapes}

\abbreviations{LLE,MD,Metadynamics}
\keywords{American Chemical Society, \LaTeX}

\begin{document}

\begin{tocentry}
\centering
\includegraphics[height=4.0cm]{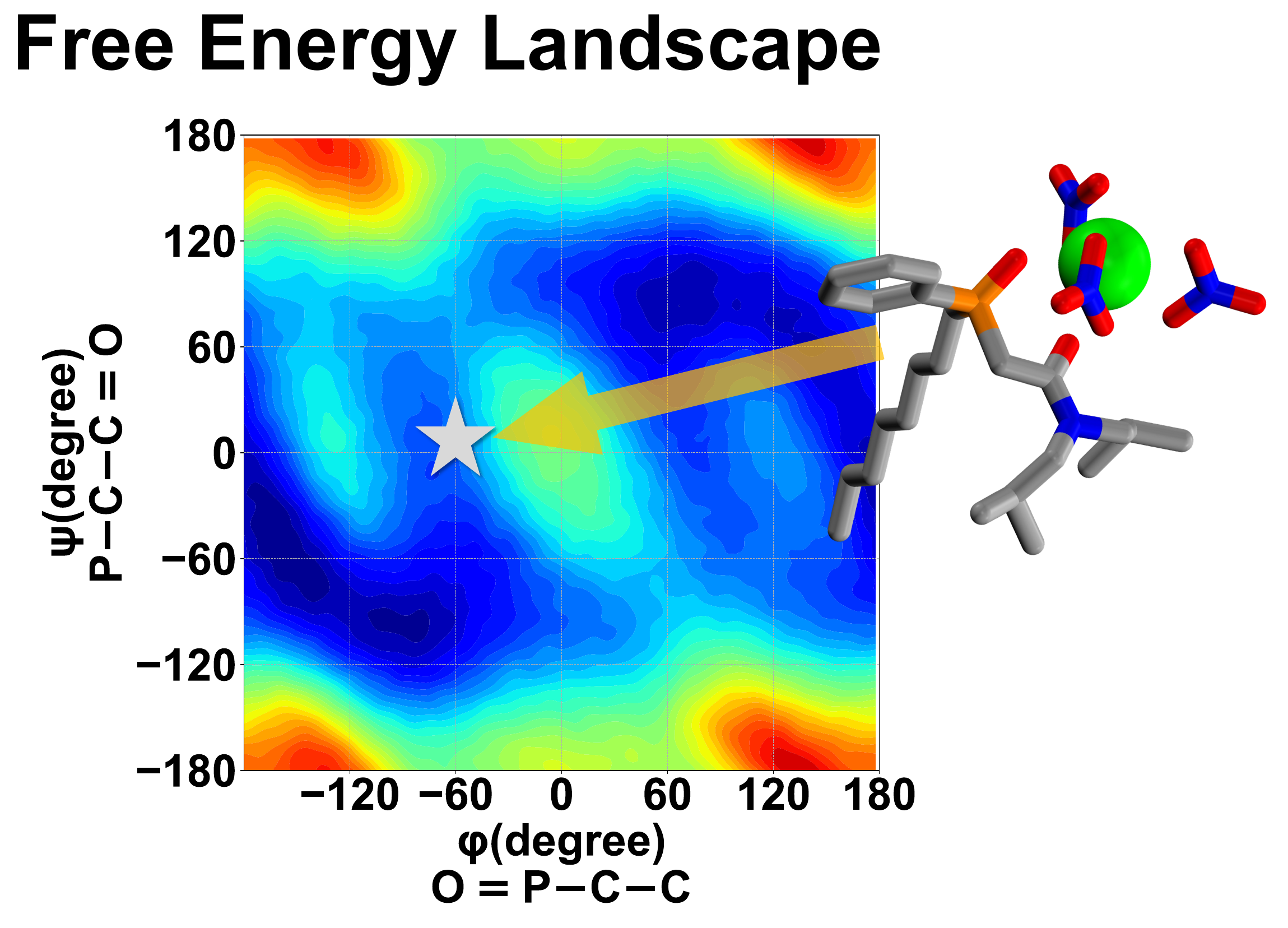}

\end{tocentry}

\newpage
\begin{abstract}

Understanding the impact of extractant functionalization on metal binding energetics in liquid-liquid extraction (LLE) is essential to guide the development of better separations processes.
Traditionally, computational extractant design uses electronic structure calculations on metal-ligand clusters to determine the metal binding energy of the lowest energy state.
Although highly accurate, this approach does not account for all the relevant physics encountered under experimental conditions.
Such methodologies often neglect entropic contributions, such as temperature effects and ligand flexibility, in addition to approximating solvent-extractant interactions with implicit solvent models.
In this study, we use classical molecular dynamics simulations with an advanced sampling method, metadynamics, to map out extractant molecule conformational free energies in the condensed phase.
We generate the complete conformational landscape in solution for a family of bidentate malonamide-based extractants with different functionalizations of the head group and the side chains.
In particular, we show how such alkyl functionalization reshapes the free energy landscape, affecting the free energy penalty of organizing the extractant into the \textit{cis}-like metal binding conformation from the \textit{trans}-like conformation of the free extractant in solution.
Specifically, functionalizing alkyl tails to the center of the head group has a greater influence on increasing molecular rigidity and disfavoring the binding conformation than functionalizing side chains.
These findings are consistent with trends in metal binding energetics based on experimentally reported distribution ratios.
We also consider a different bidentate extractant molecule, CMPO, and show how the choice of solvent can further reshape the conformational energetic landscape.
This study demonstrates the feasibility of using molecular dynamics simulations with advance sampling techniques to investigate extractant conformational energetics in solution, which, more broadly, will enable extractant design that accounts for entropic effects and explicit solvation.

\end{abstract}

\section{Introduction}

Liquid liquid extraction (LLE), also called solvent extraction, is a commonplace hydrometallurgical process used to separate and recover critical metals, such as rare earth, transuranic, and platinum group elements. \cite{hidayah2018evolution}
In LLE, metal ions are transported from an aqueous feed phase, often acidic, to an immiscible organic phase that typically contains an amphiphilic extractant molecule, which binds the targeted metal ions, and a nonpolar solvent, also called the diluent.
LLE is a thermodynamic process driven by small free energy differences in metal solvation between phases, in which large energetic terms cancel out, resulting in an overall energy on the order of $-k_{B}T$ to ensure reversibility for further reprocessing. \cite{spadina2021molecular}
While an LLE process involves many energetic terms that are often challenging to isolate and quantify, their net contributions are reflected in the overall separations performance.
Specifically, the distribution ratio is the ratio of concentration of the target metal ion in the extractant phase to the aqueous phase.
While metal-extractant binding is typically a strong, enthalpically favorable interaction, for multidentate extractants, the reorganization of the extractant into its binding conformation is often highly unfavorable. \cite{hay2004structural,boehme2002carbamoylphosphine}
Previous work using density functional theory (DFT), including by Hay and coworkers,\cite{hay2004structural} has improved separation behavior by lowering the unfavorable reorganization energy of the extracting ligand through extractant molecular design.\cite{lewis2011highly,mccann2016computer,ellis2017straining,zhang2022advancing}
In these works, a handful of molecular structures are investigated by geometry optimizations at 0 K followed by single-point energy calculations.
While capable of computing highly accurate energies, such a strategy is insufficient to provide a full picture of molecular energetics at temperature in the solution phase, as the computational cost of DFT calculations prohibits consideration of the ensemble of structures potentially relevant to metal extraction under experimentally realistic conditions.
Moreover, since geometry optimization is carried out at 0 K, neither the flexibility of molecular structure nor the contribution of entropy are considered.
In addition, solvents are typically treated implicitly by mean-field theories, which neglect that, at the molecular level, solvents can interact with extractants in more complex ways. 

Herein, we present classical molecular dynamics (CMD) simulations and profile the full energetic landscape of extractant conformations, with solvent molecules being modeled explicitly.
To enhance the sampling of all extractant conformations, especially for unfavorable high-energetic ones, a free energy estimation technique, metadynamics (MTD) \cite{laio2002escaping,barducci2011metadynamics,valsson2016enhancing}, is carried out by adding repulsive bias potentials as functions of observable collective variables. 
Unlike regular unbiased CMD, these time-evolving bias potentials discourage the system from visiting low-energetic configurations, which provides a way of mapping out the whole free-energy surface defined by the collective variables with reasonable computational cost.
In contrast to DFT techniques, which necessitate the simplification of solvents to avoid expensive calculations, CMD uses semiempirical force fields to explicitly account for solvent molecules.
While empirical CMD potentials are less accurate than those calculated from DFT, we are able to benchmark the force field performance against DFT, and then explore the full range of conformations present in the solution phase.
By using DFT to identify the binding geometry of the extractant molecule in the presence of the metal, we can then compute the free energy cost of the molecule organizing into that conformation.
In a practical system, this energetic cost competes with the favorable metal-extractant binding.
Reducing this unfavorable extractant conformational energy penalty through extractant pre-organization is an established strategy in ligand design. \cite{hay2004structural,mccann2016computer,lumetta2002deliberate,merrill2011unusual,merrill2011metal,gephart2009complexation,williams2009strong,lashley2016highly}
Here, by using MTD with CMD, we seek to extend this approach using a methodology that can apply to realistic, solution-phase systems.
By isolating this conformation energy in the absence of the metal, but at temperature and in the presence of explicit solvent, we can isolate how changes in the extractant structure reshape the conformational energetic landscape.
This will quantify how the extractant structure favors or disfavors the binding conformation compared to the lowest energy conformation that is typically assumed by the nonbinding extractant.
Overall, this will provide a complete picture of the inherent conformational energetics of an extractant molecule, allowing us to connect changes to this landscape to separations performance.

\begin{figure}
    \centering
    \includegraphics[width=1.0\linewidth]{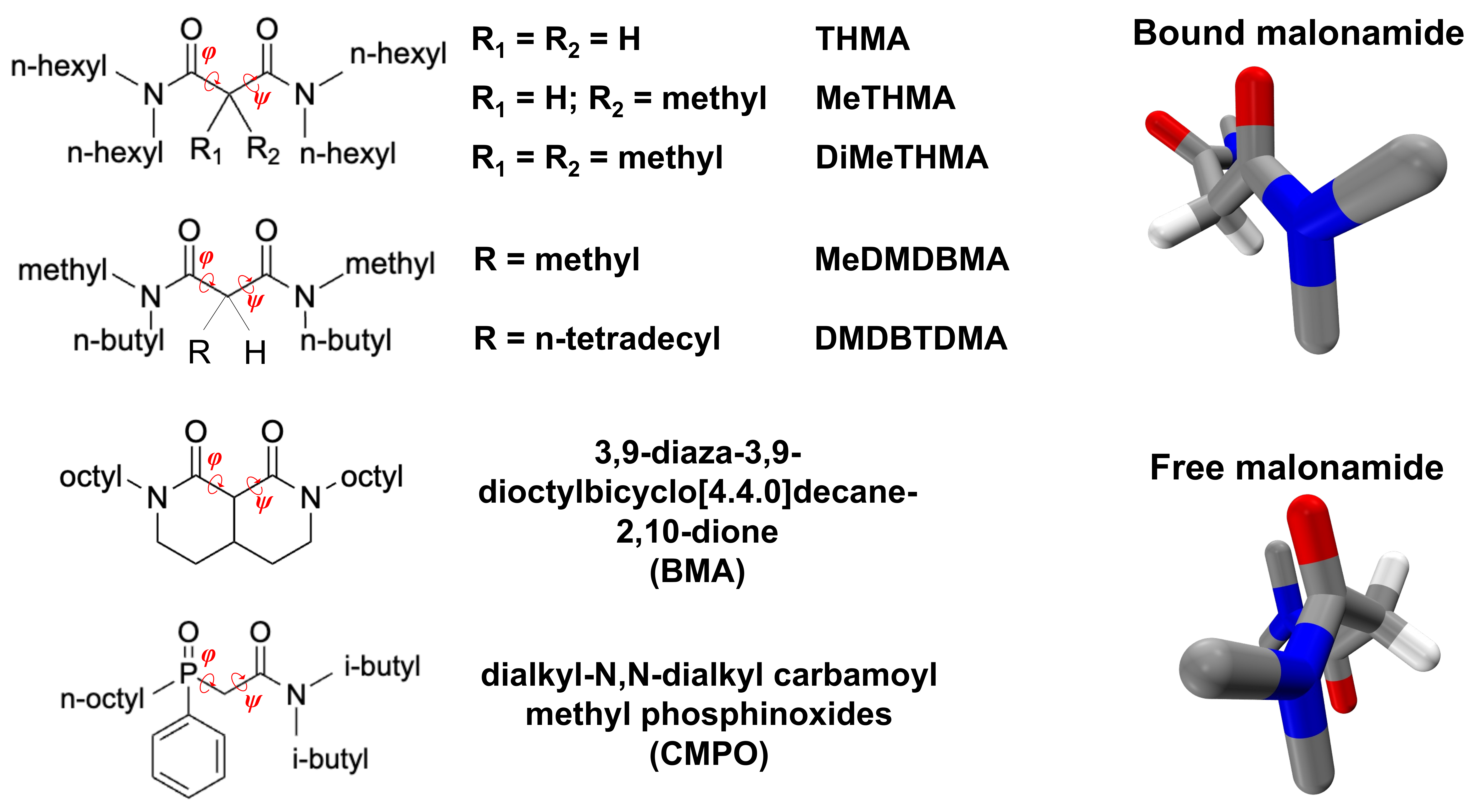}
    \caption{Left panel: schematic molecular structures of extractants considered in this study. Right panel: 3D schematic molecular structures of a bound and a free malonamide. Terminal carbons and hydrogens are omitted for simplicity.}
    \label{fig:molecules}
\end{figure}

In this work, we consider two major types of extractants: malonamides and carbamoylmethylphosphine oxides (CMPO).
For the malonamide type, we choose $N,N,N',N'$-tetrahexyl malonamide (THMA), $N,N,N',N'$-tetrahexyl-2-methyl malonamide (MeTHMA), $N,N,N',N'$-tetrahexyl-2,2-dimethyl malonamide (DiMeTHMA), $N,N'$-dimethyl-$N,N'$-dibutyl-2-tetradecyl malonamide (DMDBTDMA), and 3,9-diaza-3,9-dioctylbicyclo[4.4.0]decane-2,10-dione (BMA), which have shown significantly different separation performances with slight modifications in molecular structures. \cite{mcnamara1999extraction,lumetta1999complexation,nigond1995recent,lumetta2002deliberate}
For the type of CMPO, we select the most famous derivative, n-octyl(phenyl)-$N,N'$-diisobutyl CMPO (we will simply refer to it as CMPO in the following context), which has been implemented in the \underline{TR}ans\underline{U}ranic \underline{EX}traction (TRUEX) process. \cite{philip1985truex}
\cref{fig:molecules} shows the molecular structures of these extractants covered in this work.
By constructing the two adjacent torsion angles in the extractants' head group as collective variables (shown as $\psi$ and $\phi$ in \cref{fig:molecules}), we are able to sample and generate the full conformational free-energy landscapes by varying $\psi$ and $\phi$.
We also show 3D schematic molecular structures of bound and free extractants revealed by DFT calculations, in which the two carbonyl groups in the head groups point in different directions. It is worth mentioning that the two carbonyl groups are separated by the $\alpha$-carbon, which rationalizes our choice on collective variables of using $\psi$ and $\phi$, rather than the out-of-plane angle of the two carbonyl groups. Zero angle of $\psi$ and $\phi$ is defined as when the dihedral is in the \textit{cis} conformation.
Such free energy landscapes are able to represent full energetic information of these multidentate extractant conformations, including the free unbound \textit{trans}, transitional \textit{gauche} and bound \textit{cis} conformations.
The dramatically different energetic landscapes across different extractant functionalizations provides different perspectives on how LLE separations can be influenced by the molecular structure and conformational energetics.

\section{Simulation methodology}

\subsection{DFT calculations}

For all DFT calculations, we use a complex consisting of one Eu(III), with one bidentate malonamide or CMPO and three bindentate nitrate anions to fill the coordination vacancies and neutralize the complex.
The Eu(III) cation has 7 -- 9 coordination sites, and we choose 8 as the coordination number in all DFT calculations for simplicity.
Note that the coordination chemistry of lanthanides and extraction stoichemetry are beyond the scope of this study; the metal-extractant complexes are selected only to provide a preliminary understanding of the extractant geometry as it binds with the metal ion, without deliberately considering variation in stoichimetry of the complex or nitrate denticity.
The geometry of the Eu(III)-extractant complex is optimized with M06 \cite{zhao2008m06} hybrid functional theory.
The Stuttgart/Dresden relativistic small-core (RSC) effective core potential (ECP) basis set, obtained from Basis Set Exchange \cite{pritchard2019new}, is used for Eu(III), while 6-31G(d) is used for other elements in the complex.
All DFT calculations are performed in Gaussian 16 \cite{g16}.

\subsection{CMD and MTD}

CMD uses semi-empirical force fields to describe the interaction between bonded and nonbonded atoms in the simulation system.
By solving the Newtonian equation of motion, atomic positions are propagated with time, from which ensemble quantities are averaged and compared with experiments.
One of the challenges of CMD is that the simulated system is often ``trapped'' in local minima on the free energy surface (FES), which prevents the CMD from exploring the entire configurational space.
The detailed underlying theory of how MTD works is beyond the scope of this work; instead, we will only briefly introduce the concept of MTD.
MTD is a type of advanced sampling method that operates by spawning history-dependent Gaussian bias potentials to the original FES, which discourages the system from re-visiting low-energetic configurations and eventually helps to ``escape'' local energy minima.
The MTD simulation will stop when the whole configurational space is explored, and the negative of the summation of all spawned Gaussian bias potentials will be removed from the original FES.

For all simulations, we model the extractants using the \underline{G}eneral \underline{A}MBER \underline{F}orce \underline{F}ield (GAFF2) \cite{he2020fast} with partial charges using the Austin Model 1 with bond charge correction (AM1-BCC) method. \cite{jakalian2000fast,jakalian2002fast}
We also test the Optimized Potentials for Liquid Simulations (OPLS) force field \cite{jorgensen1996development} (see below).
The parameter set for dodecane is taken from an optimized GAFF model to prevent unphysical freezing at room temperature. \cite{vo2015computational} 
The Lorentz-Berthelot mixing rule is used for all Lennard-Jones cross-term interactions.
A \textit{trans} extractant molecule is packed with 150 dodecane molecules using PACKMOL \cite{martinez2009packmol}.
The simulation box volumes are equilibrated for 1 ns in NPT ensembles with a Nose-Hoover style thermostat and barostat \cite{martyna1994constant,parrinello1981polymorphic,shinoda2004rapid} at 303 K and 1 atmosphere.
Then, simulations are switched to NVT ensemble with Nose-Hoover style thermostat at 303 K for 150 ns MTD sampling.
All simulations are carried out with a 1 fs timestep.
For MTD simulations, every 1000 steps, a Gaussian bias potential is added to the dihedral angles $\psi$ and $\phi$ defined in \cref{fig:molecules}.
The rate of spawning the bias potential should be slower than the relaxation of the amphiphilic tails of the extractant in solution. For the rate we used in this work, \citeauthor{dasari2020conformational} has reported that it worked for a similar system in ionic liquids. \cite{dasari2020conformational} In addition to that, we cross-check that this hyperparameter can produce conformational free energy landscapes which satisfy the assumed symmetry of the malonamide extractants due to the definition of $\phi$ and $\psi$. More details on the symmetry of extractant molecules are discussed in the next section.
The height of the potential is 0.5 kJ/mol, and the width is 0.1 rad for both $\psi$ and $\phi$.
Convergence tests have been carried out to ensure that the 150 ns is sufficient to provide energetic information, especially for low energy areas with $\Delta F <$ 50 kJ/mol.
These results are shown in Figures 1S to 14S in SI.
We use LAMMPS \cite{thompson2022lammps} as the CMD engine, and MTD simulations are performed with the open source, community-developed PLUMED software package. \cite{tribello2014plumed}
Input scripts for LAMMPS and PLUMED are provided through Zenodo: http://10.5281/zenodo.10127153.

\section{Results and discussions}

\subsection{Force field comparison}

\begin{figure}[h]
    \centering
    \includegraphics[width=1.0\linewidth]{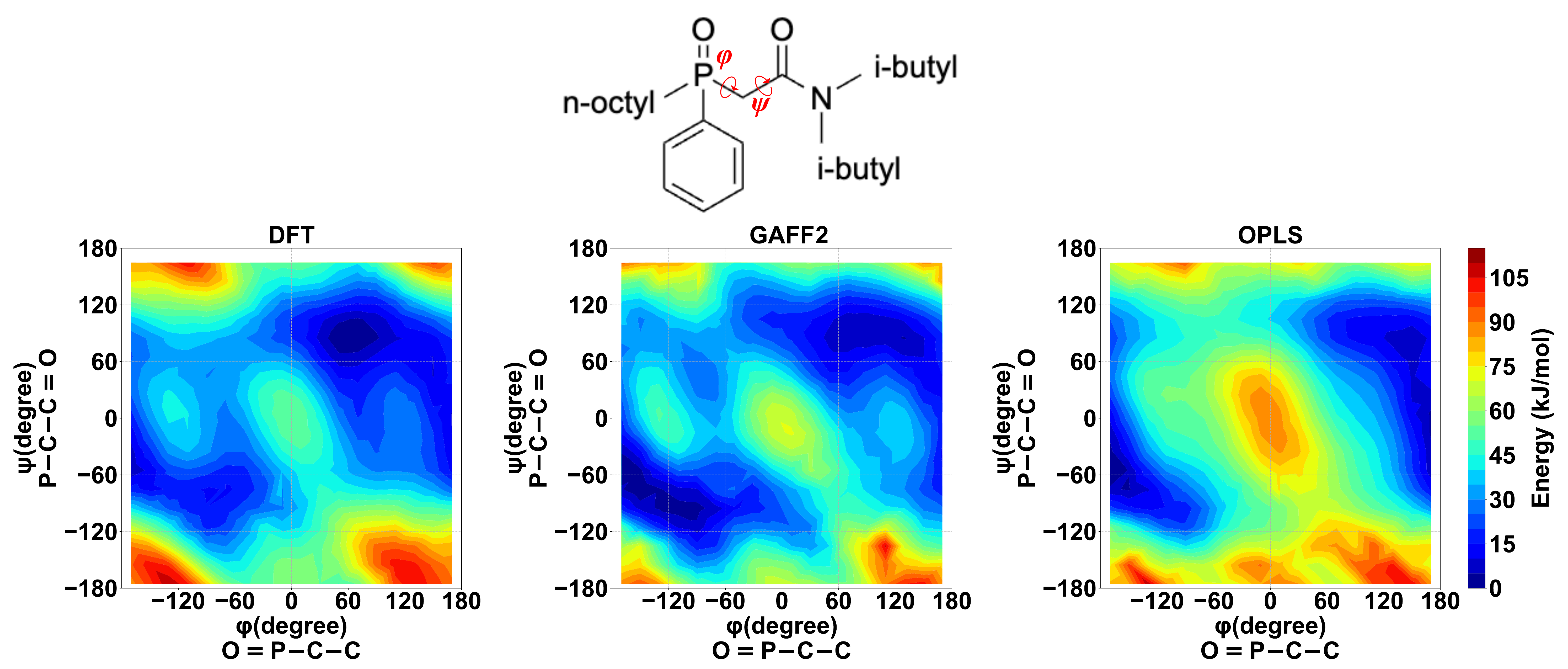}
	\caption{PESs of CMPO in gas-phase DFT calculations (left), GAFF2 (middle), and OPLS (right). Axes are defined by the corresponding dihedral angles in \cref{fig:molecules}. All energy contour maps are sharing one color-coding in the right panel.}
    \label{fig:ff}
\end{figure}

To evaluate whether the general purposed classical MD force fields are able to capture the conformation energetics of the extractant molecules accurately, we compare a single gas-phase molecule modeled using GAFF2 and OPLS force fields to DFT calculations.
We consider the CMPO extractant because the molecule is not symmetric and has a phosphine oxide group which may be poorly defined in these general-purposed classical force fields.
\cref{fig:ff} shows the Ramachandran plots for molecular mechanics and quantum mechanics.
The Ramamchandran plot \cite{RAMACHANDRAN196395} is a 2D heat map showing the properties of interest in different combinations of two adjacent dihedrals: here, we use it to show the potential energy surfaces (PES) by varying the dihedral angles of \ce{O=P-C-C} ($\phi$) and \ce{P-C-C=O} ($\psi$) in CMPO (shown in \cref{fig:molecules}), which control the head group conformations of the extractant.
The PES for GAFF2 shows that, compared to DFT calculations, the classical force field is able to capture all the fine features, given that we have not performed any reparameterizations for the CMPO molecule.
Major deviations from DFT in GAFF2 are observed only in some high-energy areas at extreme values of $\phi$ and $\psi$.
Such deviations are acceptable considering that classical force fields, which use harmonic bond and angle potentials, often miss-treat intramolecular potentials when molecules are severely distorted.
Compared to GAFF2 and DFT, OPLS does not accurately describe molecular mechanics: OPLS captures neither the shape nor the peak height of many features of the PES.
Therefore, we proceed with the CMD simulations using GAFF2.

\subsection{Effects of alkyl substitution on malonamide $\alpha$-carbon}

To test the impact of functionalization on the head group of the extractant, we investigate three malonamides where the substitutions of $\alpha$-carbon are varied. 
The addition of alkyl functional groups to the $\alpha$-carbon that joins the amide groups in the malonamide molecule is known to affect separations performance. \cite{lumetta1999complexation,mcnamara1999extraction}
Specifically, when MeTHMA and THMA are compared, the middle methyl group on the head group is known to result in lower distribution ratios for both uranyl and Eu(III) extraction.\cite{lumetta1999complexation,mcnamara1999extraction}
A more significant drop in the distribution ratios for DiMeTHMA has been observed by grafting the second methyl group, where nearly negligible extraction was reported for both uranyl and Eu(III) metal ions. \cite{lumetta1999complexation,mcnamara1999extraction}
We expect that this effect will be reflected in the conformational energetics of the molecule, which we explore through DFT and CMD supplemented with MTD.
In particular, we hypothesize that such drops in distribution ratios can potentially be explained by the increased free energy penalty for assuming the binding conformation with sequential $\alpha$-carbon functionalization based on comparing DFT calculations of the bound-state molecular geometry to the MTD-determined conformational free energy landscape.

\begin{figure}[h]
    \centering
    \includegraphics[width=1.0\linewidth]{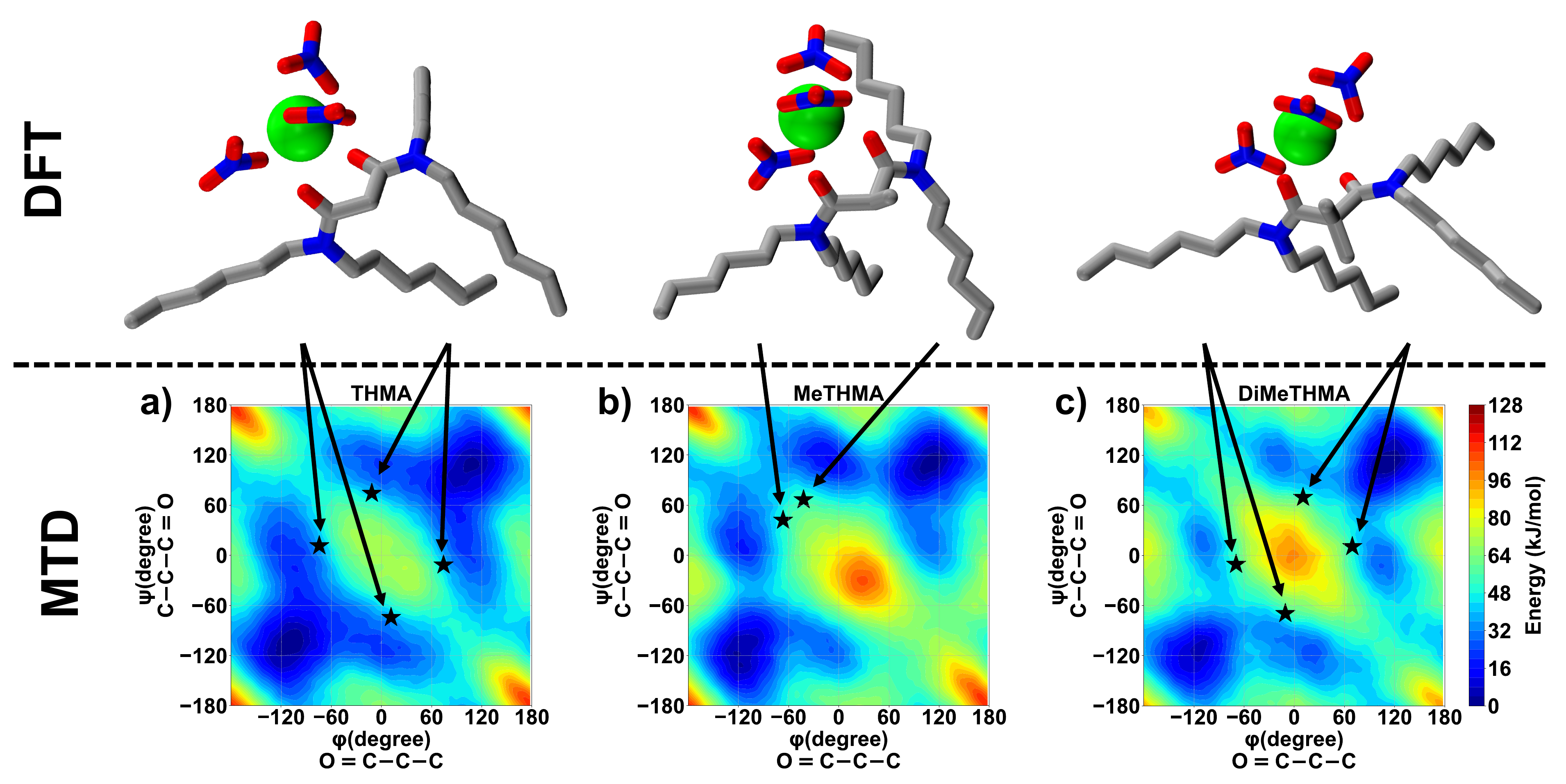}
	\caption{Top: bound state structures optimized by DFT. Bottom: Ramachandran plots of a) THMA, b) MeTHMA, c) DiMeTHMA in dodecane from MTD simulations. Axes are defined by the corresponding dihedral angles in \cref{fig:molecules}. All energy contour maps are sharing one color-coding given in panel c). Stars in each map represent the DFT-optimized conformations shown at the top, when extractants bound with an Eu(III) ion.}
    \label{fig:MA}
\end{figure}

In order to identify the bound-state extractant conformation, we first carry out DFT calculations in which an extractant molecule forms a complexant with one Eu(III) and three nitrates.
The molecular structures of these clusters are shown in the top panels of \cref{fig:MA}.
As we discussed above, compared with static DFT calculations, CMD can provide more details on the extractant confomrational energetics, especially when temperature and solvent effect are taken into consideration.
The bottom panels in \cref{fig:MA} show Ramamchandran plots, in which the conformational free energy landscapes are mapped from CMD simulations using the MTD sampling technique.
We note that, unlike the potential energy surface calculated from single gas-phase configurations described in the force field validation and shown in \cref{fig:ff}, the MTD simulations provide free energy surfaces.
The optimized DFT structures of the extractant bound with the Eu(III) provide the combinations of $\phi$ and $\psi$ assumed to define the binding state, which we project onto the 2D Ramamchandran free energy surface to show where the binding conformation falls in the total energetic landscape of the extractant.
We calculate one DFT structure per malonamide; the multiple bound states in \cref{fig:MA} are inferred from molecular symmetry.
Note that our DFT calculations are only used to determine $\phi$ and $\psi$ at the bound state conformation of a single extractant, and we are not trying to replicate any particular experimental stoichiometry of the metal-extractant complex.
Nevertheless, all our DFT optimized structures suggest that \textit{cis}-malonamide binds with the metal ion in a bidentate fashion, consistent with experimental studies. \cite{ellis2012coordination,ellis2013periodic,ellis2014complexation}

With the binding conformation of each malonamide identified, we proceed with CMD using MTD to determine how changes to the molecular structure affects the \textit{cis} conformation in the absence of the metal.
From inspection of the binding geometries and conformational free energy surface, the fundamental competition between metal binding and molecular deformation is apparent.
The enthalpically favorable metal binding process favors the \textit{cis} conformation, while the intramolecular conformational energetics prefer \textit{gauche} and, even more so, \textit{trans} states.
Therefore, changes in the extractant that move the accessible \textit{cis}-like regions further from the true \textit{cis} region near $\phi = 0$ and $\psi = 0$ presumably weaken the metal-extractant binding.
Thus, when the conformational energetic penalty for metal binding is added to the favorable binding process, the overall free energy of extraction decreases.

As we have discussed above, stoichiometry is beyond the scope of this work: our DFT calculations are used to show the rough locations of the bound-state structures in the 2D Ramachandran plot rather than to reveal the exact metal-extractant complex. In contrast to the 1:1 metal:ligand stoichiometry calculated by DFT, the stoichiometries in solution are usually 2:1 and 3:1. We can expect that the steric and electronic environment in the coordination sphere of the metal ion will slightly influence the conformation of the ligand adopted. To test the effect of the stoichiometry, we have performed a DFT calculation on a cluster containing 1:2 Eu:THMA. The DFT calculation has shown that $\phi$s and $\psi$s for the two THMA in the Eu:THMA=1:2 cluster are (-38$^{\circ}$,50$^{\circ}$) and (-31$^{\circ}$,44$^{\circ}$). Compared to the stars at (-12$^{\circ}$,74$^{\circ}$), (12$^{\circ}$,-74$^{\circ}$), (74$^{\circ}$,-12$^{\circ}$), and (-74$^{\circ}$,12$^{\circ}$) in \cref{fig:MA}, the bound-state structures in 1:2 Eu:THMA have higher free energies (65 kJ/mol vs. 38 kJ/mol). Such an observation indicates that extractant bound-state energetics are affected by the stoichiometry, and this energetic penalty is presumably counterbalanced by the strong enthalpic metal-extractant binding of the additional extractant ligand. The difference between the 1:1 and 1:2 complex suggests a competition between the addition of more extractants (favorable for solvating) and the increase in conformational energetic cost (unfavorable). This difference also demonstrates the importance of considering the ligand design holistically, i.e., in conjunction with the entire ensemble of ligand-metal complexes. Extractant conformations and complex stoichiometries would presumably also be affected by other aspects of the extraction system, including lanthanide size \cite{kravchuk2023structural} and counteranion identity. However, we want to reiterate that, in this work, we are focusing on the intrinsic conformational free-energy landscape of the extractant, rather than the intermolecular steric effects induced by the stoichiometry.

When a methyl group is grafted on the $\alpha$-carbon from THMA to MeTHMA, free energy Ramachandran plots have shown that the flexibility of the extractant decreases: low-energetic areas are much more confined for MeTHMA, showing conformations that become inaccessible.
The free energy of the $\phi = 0$ and $\psi = 0$ region increases significantly for MeTHMA, making the binding conformations less accessible, which potentially explains the observation that the distribution ratio drops from THMA to MeTHMA.
For THMA, since there is additional symmetry around the $\alpha$-carbon, they do not possess chirality and should also be symmetric across $y=x$ as well.
For MeTHMA, the middle alkyl group gives chirality to the molecule, so we do not expect this symmetry.
(In fact, this symmetry helps to demonstrate when the MTD trajectory has adequately sampled the conformational landscape, as the features and energy scales should be consistent across the applicable symmetry planes.)
The unfavorable \textit{cis} region also shifts to the bottom right of the Ramachandran plot.
However, the binding conformation responds to the higher energetic cost of the \textit{cis} conformation and the asymmetry of the free energy surface across $y=x$.
The optimized bound state $\phi$ and $\psi$ identified by DFT shifts to the upper left corner of the \textit{cis} region for MeTHMA compared to THMA, reflecting these changes to the underlying free energy surface.
The DFT optimized bound-state structures also show that although the \textit{cis} conformation is most relevant for both THMA and MeTHMA, these structures are slightly different, where carbonyl groups are more parallel for MeTHMA.

By adding the second methyl group on the $\alpha$-carbon, going from MeTHMA to DiMeTHMA, the most significant change in \cref{fig:MA} is that the free energy landscape becomes symmetric across $y=x$ again.
The DFT optimized bound state $\phi$ and $\psi$ also shift back to positions more similar to THMA.
Overall, compared with the other malonamides, the low-energetic areas in the free-energy landscape become even more confined.
The central (\textit{cis}) high-energy area shifts back closer to $\phi = 0$ and $\psi = 0$, resulting in the disappearance of a low-energy area that is easily accessible for the bound state.
These effects potentially explain why DiMeTHMA only shows negligible extraction performance for Eu(III) and uranyl metal ions, where the high energetic barrier near the \textit{cis} conformation makes the binding state for DiMeTHMA much less energetically accessible.

\subsection{Effects of alkyl tail length on malonamides}

\begin{figure}[h]
    \centering
    \includegraphics[width=1.0\linewidth]{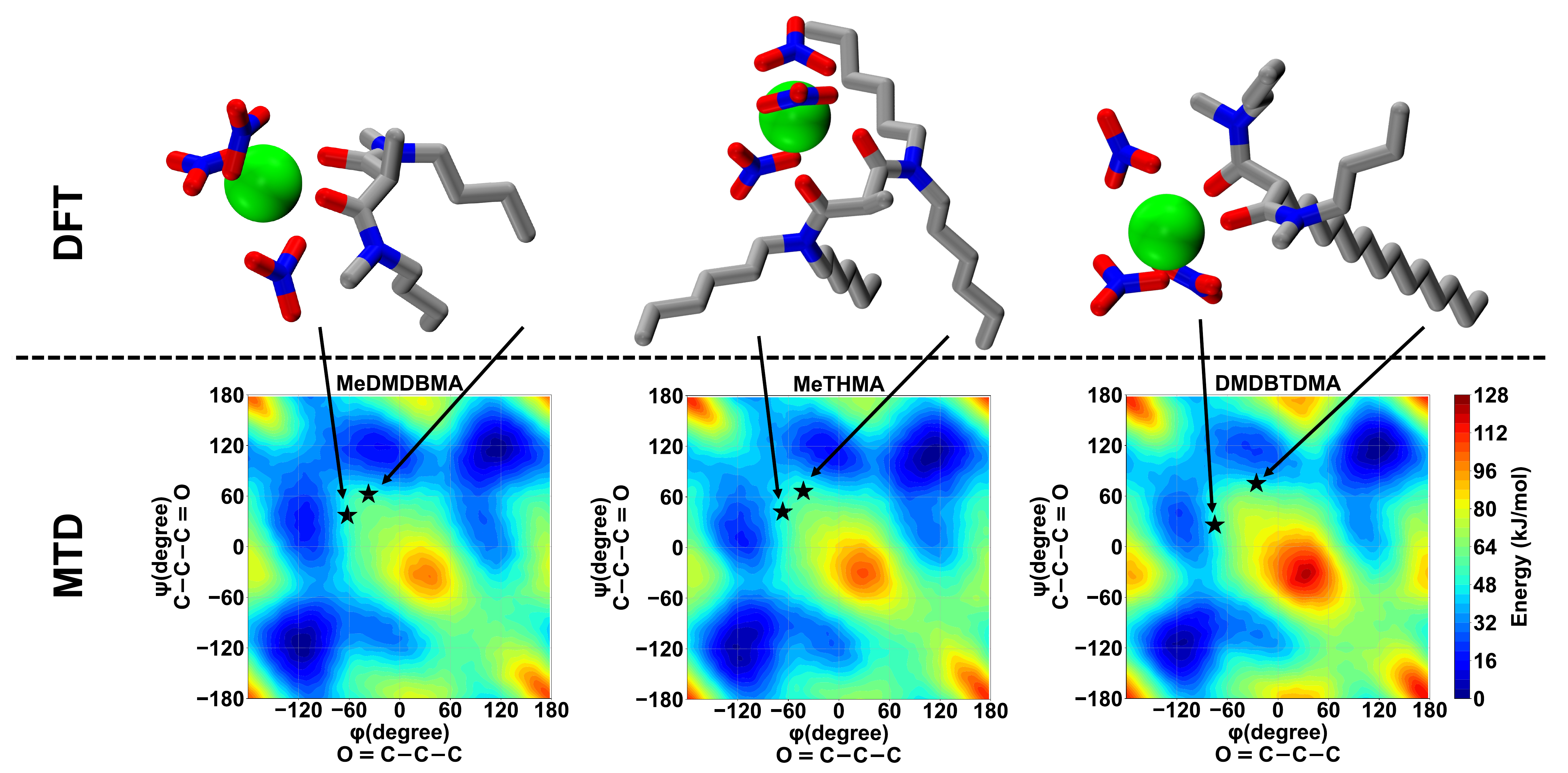}
	\caption{Top: bound state DMDBTDMA optimized by DFT; bottom: Ramachandran plot of DMDBTDMA in dodecane from MTD simulations. X and Y axes are defined by the dihedral angles in \cref{fig:molecules}. Stars represent the DFT-optimized conformations shown at the top, when the extractant bound with an Eu(III) ion.}
    \label{fig:DMDBTDMA}
\end{figure}

The functionalization of longer alkyl chains provides better solubility in the organic phase,\cite{berthon2007solvent,bonnett2023critical} improving the metal loading capacity. 
However, the potential effect of alkyl length on conformational flexibility is not understood.
To test the effect of side chain lengths, we first compare MeDMDBMA, an analogue of MeTHMA but with shorter and different alkyl side chains.
The left and center panels in \cref{fig:DMDBTDMA} shows the free energy landscapes of MeDMDBMA and MeTHMA, respectively.
The free energy maps of these two molecules are similar, with DFT-determined bound-state conformations at similar locations on the Ramachandran plot.
We also consider DMDBTDMA, an extractant used in the DIAMEX (DIAMide Extraction) process for the recovery of trivalent actinide ions,\cite{erlinger1998effect} which has a much longer alkyl tail, replacing the methyl group on the $\alpha$-carbon of MeDMDBMA.
Noting that, compared to THMA, slightly better separation performance has been reported with DMDBTDMA for uranium, \cite{makombe2022uranium} it is worth mentioning that other interaction contributions, such as the assembly of multiple extractants and the complexation with metal ions, will also play important roles in driving separation.
It is likely that the additional sides chains of THMA introduce steric hindrance that disfavors formation of extracted metal complexes, which we are not considering in this study.
Further simulation studies using MTD on metal-extractants solvates could identify differences in steric effects between extractants with different side chain lengths.
Overall, the unfavroable regions for DMDBTDMA have energies higher than those of MeTHMA and MeDMDBMA, indicating that the length of the alkyl group on the $\alpha$-carbon has a greater influence on increasing conformational energetics and lowering the head group's flexibility than on the side chain.
While the addition of an alkyl tail length to the molecule is necessary to improve its solubility, these results suggest that different locations on the malonamide affect its conformation flexibility and, presumably, its separation performance.

\subsection{Bicyclic malonamide}

\begin{figure}[ht]
    \centering
    \includegraphics[width=0.6\linewidth]{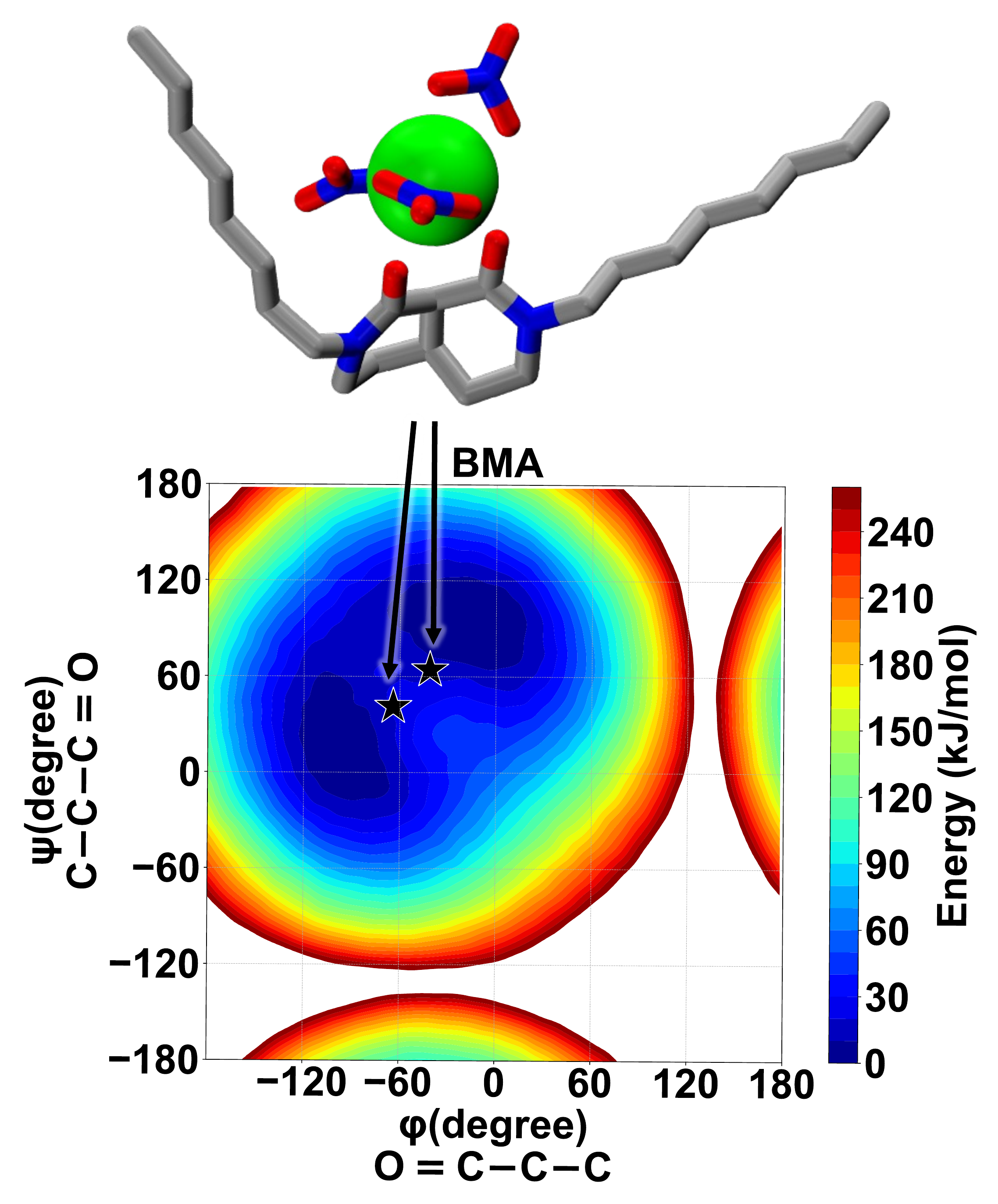}
	\caption{Top: bound state BMA optimized by DFT; bottom: Ramachandran plot of BMA in dodecane from MTD simulations. X and Y axes are defined by the dihedral angles in \cref{fig:molecules}. Stars represent the DFT-optimized conformations shown at the top, when the extractant bound with an Eu(III) ion. We use the white color to represent the region, where the CMD simulation cannot sample even with MTD.}
    \label{fig:BMA}
\end{figure}

A common strategy in ligand design to improve selectivity and separation performance is to rigidify the backbone of the molecule, which preorganizes the molecule into the preferred binding conformation. \cite{lewis2011highly,johnson2023size}
For example, to design a better malonamide for rare earth separation, \citeauthor{lumetta2002deliberate} synthesized a bicyclic malonamide, finding 7 orders of magnitude improvement in the distribution ratio of Eu(III).\cite{lumetta2002deliberate}
Such a significant increase in the distribution ratio was achieved through restricting the head group to the binding conformation by introducing a bicyclic carbon group.
We use the MTD to map out the free energy landscape of this bicyclic malonamide, with the Ramachandran plot shown in \cref{fig:BMA}.

The conformational free-energy landscape of the bicyclic malonamide is dramatically different from that of all of the malonamides investigated above.
The \textit{trans} conformation, typically where the global mininum resides, completely disappears.
Instead, the extrantant molecule is now restricted in the binding, \textit{cis} conformation, shown in blue colors in the top left quadrant.
There is a wide region (colored by white) where our CMD cannot explore this configurational space even with the MTD, due to the high-energy barrier.
The bound-state structures calculated with DFT are much closer to the minima than with the other malonamides.
The energetic difference between the bound structure and the minimum free energy conformation is around 20 kJ/mol, which is significantly less than those of the malonamides investigated above, although not sufficiently less to explain the seven orders of magnitude increase in the Eu(III) distribution compared to THMA reported by \citeauthor{lumetta2002deliberate}.

\subsection{CMPO: Asymmetric binding moieties}

\begin{figure}[H]
    \centering
    \includegraphics[width=1.0\linewidth]{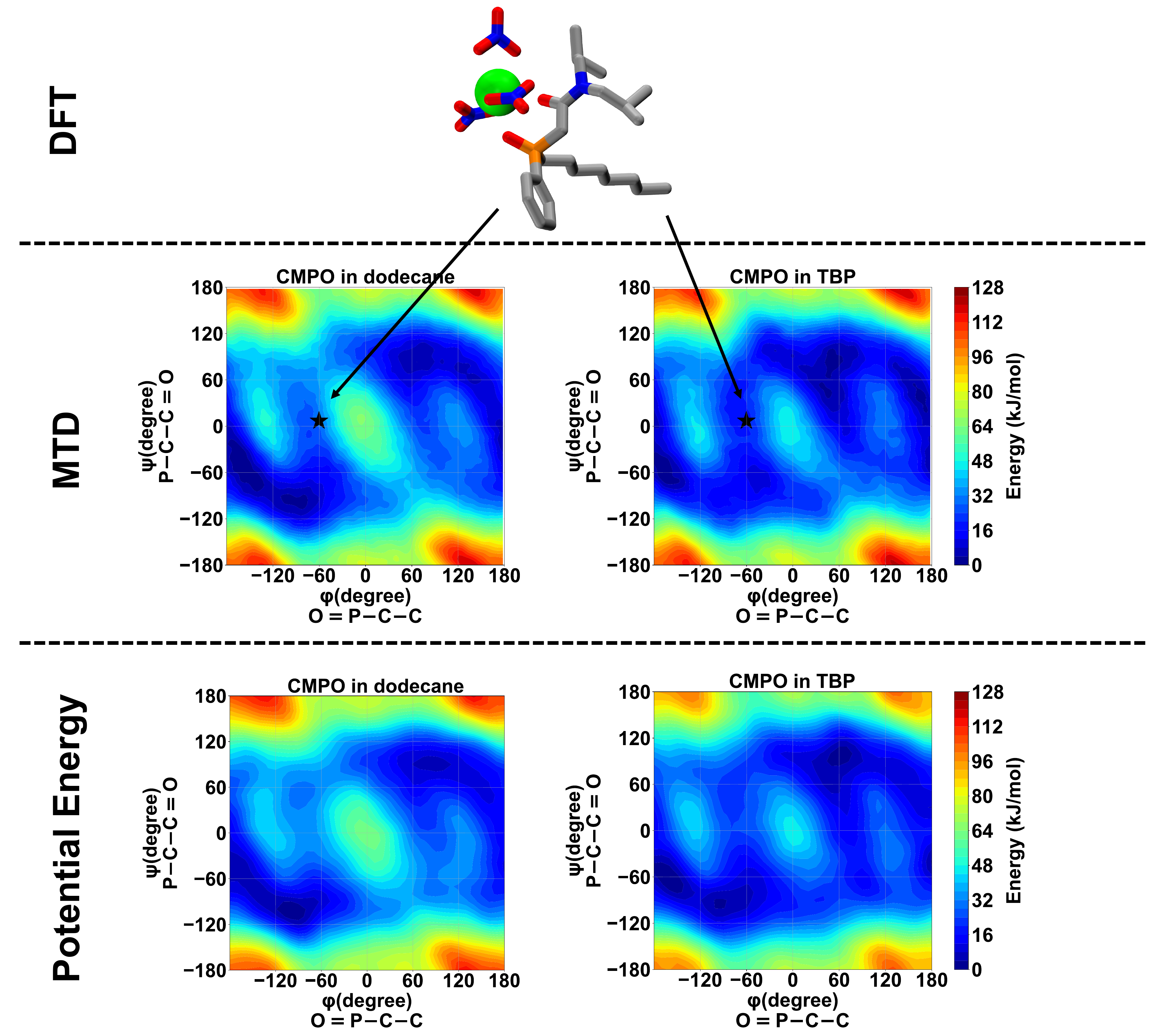}
	\caption{Top: bound state CMPO optimized by DFT; Middle: Ramachandran plot of CMPO in dodecane and TBP from MTD simulations; Bottom: Ramachandran plot of CMPO's potential energies sampled from MTD trajectories. X and Y axes are defined by the dihedral angles in \cref{fig:molecules}. Stars represent the DFT-optimized conformations shown at the top, when the extractant bound with an Eu(III) ion.}
    \label{fig:CMPO}
\end{figure}

Compared with traditional PUREX (Plutonium Uranium Reduction EXtraction), CMPO was designed to improve the functionality and separation of actinides in LLE. \cite{kalina1981extraction,horwitz1982selected,philip1985truex}
Compared to the malonamides that we have covered above, CMPO not only has a highly asymmetric molecular structure, but also has an entirely different functional head group consisting of the phosphine oxide group in addition to the carbonyl group.
Here, we will use CMD with MTD to investigate its conformational behavior and show its differences in comparison to that of the malonamides.
The left panel in \cref{fig:CMPO} shows the energetic landscape of CMPO in dodecane, with its asymmetry readily apparent in the conformation free energy surface.
In comparison to the malonamides we investigated, the conformational free energies of CMPO are generally much shallower, not only for the bound conformation but also for other nearby conformations.
Such a lower energy landscape indicates that more conformations are easily accessible at room temperature, which gives CMPO greater flexibility in solution.
This lower overall energy of the binding state of CMPO could explain its increased performance compared with that of malonamides, in addition to the greater basicity of the phosphine oxide binding site.

A primary benefit of studying conformational energetics with CMD using MTD is the ability to include explicit solvation.
To explore this, we also investigate CMPO in a solution of TBP, which is a more polar solvent than dodecane with its dielectric constant of 8.178 compared to 2.006 for dodecane.
Our DFT calculations with the PCM implicit solvent model show that the potential energy (not free energy) of the \textit{cis} conformation ($\Delta E_{cis-trans}$) is stabilized by 9 kJ/mol in TBP compared to dodecane.
However, it is still challenging to use DFT calculations to map out the whole free energy landscape for all the possible combinations of $\phi$ and $\psi$ that define the configuration of the head group.
Our MTD simulations for CMPO in TBP (shown in the \cref{fig:CMPO} right panel) agree with the above DFT calculations in that the bound \textit{cis} conformation is more strongly stabilized by TBP compared to dodecane, with the difference in free energy between the DFT-calculated bound conformation decreasing by 15 kJ/mol going from dodecane to TBP.
The differences between this value and that obtained from DFT could result from some combination of differences in the energetic description (empirical force field versus quantum mechanical), the explicit versus implicit solvent, and being a free energy rather than a potential energy.
Another observation from the free energy surface is that not only is the central ``hot spot'' corresponding to the \textit{cis} conformation greatly lowered in TBP solvent, the overall shape of the free energy landscape is altered.
Local minima in the bottom left quadrant ($\phi=-90^{\circ}$, $\psi=-90^{\circ}$) in dodecane shift to ($\phi=-50^{\circ}$, $\psi=-60^{\circ}$) in TBP.
In addition, the local minima in the top right quadrant ($\phi=60^{\circ}$, $\psi=90^{\circ}$) expand in TBP.

To deconvolute the changes in free energy landscape into the enthalpic and entropic contributions, we rerun MTD trajectories and sample potential energies of CMPO in dodecane and TBP. The potential energy is plotted in the format of $\phi$ and $\psi$ in \cref{fig:CMPO}. By comparing the middle and bottom panels, we can observe that they are similar to each other, ascribing the changes in the free energy landscape to the enthalpic contribution.
These small shifts indicate that certain CMPO conformations are stabilized through molecular-level solvent-extractant interactions that may be difficult to capture with mean-field implicit solvent methods.
Overall, our methodology of using CMD with MTD has shown its capability in investigating not only extractant conformation energetics with different head groups but also the solvent effect.
Additionally, the conformational behavior of CMPO potentially explains its superior separation performance over those of malonamides.
For reference, all of the free energy differences between the bound state and the global minimum, as well as differences in this value between all extractants considered in this study, are summarized in \cref{unbound}.

\begin{center}
\begin{threeparttable}
\caption {Free energy differences between unbound and bound states}
\label{unbound}    
\begin{tabular}{|  >{\centering\arraybackslash}p{5cm} | >{\centering\arraybackslash}p{4cm} | >{\centering\arraybackslash}p{4cm} |} 
  \hline
  Extractants & $\Delta F$ (kJ/mol) \tnote{a,b} & $\Delta\Delta F$ to THMA (kJ/mol) \tnote{c} \\ 
  \hline
  THMA in dodecane & 38 $\pm$ 1 & --- \\
  MeTHMA in dodecane & 52 $\pm$ 3 & 15 $\pm$ 4 \\
  DiMeTHMA in dodecane & 53 $\pm$ 1 & 15 $\pm$ 2 \\
  MeDMDBMA in dodecane & 52 $\pm$ 1 & 14 $\pm$ 2 \\
  DMDBTDMA in dodecane & 47 $\pm$ 1 & 10 $\pm$ 3 \\
  CMPO in dodecane & 30 $\pm$ 1 & --7 $\pm$ 2 \\
  CMPO in TBP & 15 $\pm$ 2 & --22 $\pm$ 3 \\
  BMA in dodecane & 20 $\pm$ 1 & --18 $\pm$ 2 \\
  \hline
\end{tabular}
\begin{tablenotes}[flushleft]\footnotesize
\item[a] Values are taken from averaging the bound state energies for the last 50 ns of simulations.
\item[b] Errors are estimated by the standard deviations of the bound state energies from the last 50 ns of simulations.
\item[c] Errors are propogated by $\sigma = \sqrt{(\sigma_1)^2 + (\sigma_2)^2}$
\end{tablenotes}
\end{threeparttable}
\end{center}

\subsection{Bound-state energetics and minimum free energy pathways}

Although all energetic information is contained in the free energy surface, the identification of a minimum free energy path (MFEP) between states of interest---often local or global minima---can elucidate the reaction coordinate and transition kinetics between those states. \cite{ensing2005recipe}
Briefly, the MFEP is a simplified reaction coordinate that connects two minima of interest on a free energy surface.
In solvent extraction, when the transfer of a solute between aqueous to organic phases is slow, it can be the rate-limiting process.
It is not generally known what steps along the process of forming the extractable complex dominate this process, i.e., whether conformational interconversion is rate-limiting compared to assembly of the metal, extractant and counteranions.
However, the free energy surfaces generated here provide the opportunity to measure the interconversion kinetics of the free extractant in solution.
To investigate this, we consider the MFEP of DMDBTDMA, as it is a real working extractant in solvent extraction processes, and its free energy landscape has several local minima that are close to each other due to the functionalization of $\alpha$-carbon.

\begin{figure}[h]
    \centering
    \includegraphics[width=1.0\linewidth]{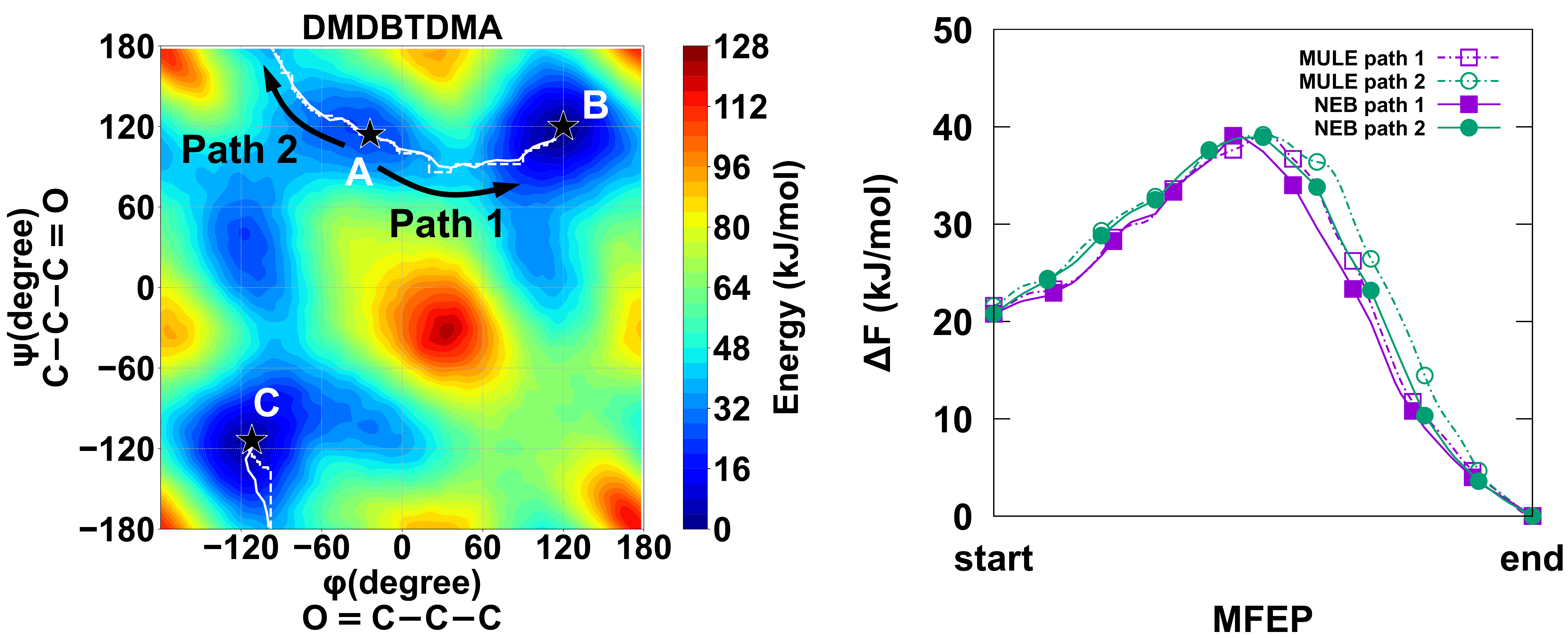}
	\caption{Left: free energies associated with the two MFEPs for DMDBTDMA. Dashed line (MULE); solid line (NEB). Right: minimum-to-minimum MFEPs identified on the free energy landscape for DMDBTDMA.}
    \label{fig:MFEP}
\end{figure}

Although the identification of MFEP is not trivial and remains research-active, here, we apply the methodology developed by \citeauthor{fu2020finding}, called MULE.
As shown in the left panel \cref{fig:MFEP}, using MULE, we identify two MFEPs from the local minimum of DMDBTDMA in basin A to the global minima in basins B and C, which are shown by dashed white lines.
Each MFEP associated with a free energy transition profile is shown on the right panel.
In order to verify the MFEP by MULE, we have identified the same minimum-to-minimum transition paths using another popular methodology, the nudged elastic band (NEB) \cite{henkelman2000improved}, shown by solid lines in \cref{fig:MFEP}.
Both NEB and MULE give similar transition paths on the free energy surface; NEB appears to do a better job on smoothing the transition path while the profiles are descending to the global minimum at basin C.
Committor analysis was performed on these two MFEPs identified by MULE, to prove that the transition paths follow the basic properties of reaction coordinates, and the results are shown in Figures 15S and 16S in the SI.
From the committor analysis, we can see a $\sim$ 50\% acceptance rate at both saddle points, lending confidence to the MFEPs identified in this work.
Overall, although path 2 goes from basin A to B through the periodic boundary of the Ramachandran plot, these two MFEPs are similar to each other in terms of shape and relative free energy of the saddle point and local minimum.
Such a multiplicity of MFEPs suggests the existence of multiple transition pathways for DMDBTDMA due to the functionalization on the $\alpha$-carbon, providing a complete picture of the molecular flexibility and kinetics of reorganization.

\section{Conclusion}

Traditional computational ligand design studies in liquid-liquid extraction of metals heavily rely on cluster DFT calculations, in which the metal-extractant binding energy is typically used as the primary means to evaluate ligand performance.
Such a strategy neglects the degrees of freedom of the extractant structure in solution in favor of a single, lowest-energy structure.
To address this limitation, we implemented a combined approach with DFT and CMD to study the impact of the extractant molecular structure on the energetic penalty associated with organizing the extractant molecule into the metal-binding conformation.
An advanced sampling method, metadynamics, enabled sampling of the entire conformational free energy landscape of a single metal-free extractant molecule in solution.
These CMD results provide entropic information, directly reporting free energies instead of potential energies, in addition to considering solvent effects explicitly.
By using DFT calculations to project bound-state structures onto the free-energy landscapes sampled from CMD, we show how functionalization on the extractants' head groups reshapes the conformational energetics.
These changes in the conformational free-energy landscape potentially explain the relationship between the functionalization of the extractant molecule and the observed separation performance as measured by the metal-ion distribution ratios.
As the unfavorable conformational free energy associated with assuming the binding conformation is added to the competing enthalpic favorability of the metal-ligand binding, changes to the conformational free energy in the absence of metal ions still provide important information on the separation behavior of the extractant.
Overall, this study highlights the opportunity to use CMD simulations with the MTD technique to explore extractant behavior, learning how to tune intramolecular conformational energetics to enhance separation performance, while also explicitly accounting for solvent-solute interactions that are present in practical LLE processes.

\begin{acknowledgement}

This work was supported by U.S. Department of Energy (DOE), Office of Science, Office of Basic Energy Sciences, Chemical Sciences, Geosciences, and Biosciences Division, Separation Science Program, under Contract DE-AC02-06CH11357 to UChicago Argonne, LLC, Operator of Argonne National Laboratory.
We gratefully acknowledge the computing resources provided on Bebop, a high-performance computing cluster operated by the Laboratory Computing Resource Center at Argonne National Laboratory.

\end{acknowledgement}

\begin{suppinfo}

Convergence test on all extractant molecules studied in this work; committor analysis for the two MFEPs identified for DMDBTDMA.

\end{suppinfo}

\bibliography{achemso-demo}

\providecommand{\latin}[1]{#1}
\makeatletter
\providecommand{\doi}
  {\begingroup\let\do\@makeother\dospecials
  \catcode`\{=1 \catcode`\}=2 \doi@aux}
\providecommand{\doi@aux}[1]{\endgroup\texttt{#1}}
\makeatother
\providecommand*\mcitethebibliography{\thebibliography}
\csname @ifundefined\endcsname{endmcitethebibliography}  {\let\endmcitethebibliography\endthebibliography}{}
\begin{mcitethebibliography}{52}
\providecommand*\natexlab[1]{#1}
\providecommand*\mciteSetBstSublistMode[1]{}
\providecommand*\mciteSetBstMaxWidthForm[2]{}
\providecommand*\mciteBstWouldAddEndPuncttrue
  {\def\EndOfBibitem{\unskip.}}
\providecommand*\mciteBstWouldAddEndPunctfalse
  {\let\EndOfBibitem\relax}
\providecommand*\mciteSetBstMidEndSepPunct[3]{}
\providecommand*\mciteSetBstSublistLabelBeginEnd[3]{}
\providecommand*\EndOfBibitem{}
\mciteSetBstSublistMode{f}
\mciteSetBstMaxWidthForm{subitem}{(\alph{mcitesubitemcount})}
\mciteSetBstSublistLabelBeginEnd
  {\mcitemaxwidthsubitemform\space}
  {\relax}
  {\relax}

\bibitem[Hidayah and Abidin(2018)Hidayah, and Abidin]{hidayah2018evolution}
Hidayah,~N.~N.; Abidin,~S.~Z. The evolution of mineral processing in extraction of rare earth elements using liquid-liquid extraction: A review. \emph{Minerals Engineering} \textbf{2018}, \emph{121}, 146--157\relax
\mciteBstWouldAddEndPuncttrue
\mciteSetBstMidEndSepPunct{\mcitedefaultmidpunct}
{\mcitedefaultendpunct}{\mcitedefaultseppunct}\relax
\EndOfBibitem
\bibitem[{\v{S}}padina \latin{et~al.}(2021){\v{S}}padina, Dufr{\^e}che, Pellet-Rostaing, Mar{\v{c}}elja, and Zemb]{spadina2021molecular}
{\v{S}}padina,~M.; Dufr{\^e}che,~J.-F.; Pellet-Rostaing,~S.; Mar{\v{c}}elja,~S.; Zemb,~T. Molecular forces in liquid--liquid extraction. \emph{Langmuir} \textbf{2021}, \emph{37}, 10637--10656\relax
\mciteBstWouldAddEndPuncttrue
\mciteSetBstMidEndSepPunct{\mcitedefaultmidpunct}
{\mcitedefaultendpunct}{\mcitedefaultseppunct}\relax
\EndOfBibitem
\bibitem[Hay \latin{et~al.}(2004)Hay, Gutowski, Dixon, Garza, Vargas, and Moyer]{hay2004structural}
Hay,~B.~P.; Gutowski,~M.; Dixon,~D.~A.; Garza,~J.; Vargas,~R.; Moyer,~B.~A. Structural criteria for the rational design of selective ligands: convergent hydrogen bonding sites for the nitrate anion. \emph{Journal of the American Chemical Society} \textbf{2004}, \emph{126}, 7925--7934\relax
\mciteBstWouldAddEndPuncttrue
\mciteSetBstMidEndSepPunct{\mcitedefaultmidpunct}
{\mcitedefaultendpunct}{\mcitedefaultseppunct}\relax
\EndOfBibitem
\bibitem[Boehme and Wipff(2002)Boehme, and Wipff]{boehme2002carbamoylphosphine}
Boehme,~C.; Wipff,~G. Carbamoylphosphine oxide complexes of trivalent lanthanide cations: Role of counterions, ligand binding mode, and protonation investigated by quantum mechanical calculations. \emph{Inorganic Chemistry} \textbf{2002}, \emph{41}, 727--737\relax
\mciteBstWouldAddEndPuncttrue
\mciteSetBstMidEndSepPunct{\mcitedefaultmidpunct}
{\mcitedefaultendpunct}{\mcitedefaultseppunct}\relax
\EndOfBibitem
\bibitem[Lewis \latin{et~al.}(2011)Lewis, Harwood, Hudson, Drew, Desreux, Vidick, Bouslimani, Modolo, Wilden, Sypula, and et~al]{lewis2011highly}
Lewis,~F.~W.; Harwood,~L.~M.; Hudson,~M.~J.; Drew,~M.~G.; Desreux,~J.~F.; Vidick,~G.; Bouslimani,~N.; Modolo,~G.; Wilden,~A.; Sypula,~M.; et~al Highly efficient separation of actinides from lanthanides by a phenanthroline-derived bis-triazine ligand. \emph{Journal of the American Chemical Society} \textbf{2011}, \emph{133}, 13093--13102\relax
\mciteBstWouldAddEndPuncttrue
\mciteSetBstMidEndSepPunct{\mcitedefaultmidpunct}
{\mcitedefaultendpunct}{\mcitedefaultseppunct}\relax
\EndOfBibitem
\bibitem[McCann \latin{et~al.}(2016)McCann, Silva, Windus, Gordon, Moyer, Bryantsev, and Hay]{mccann2016computer}
McCann,~B.~W.; Silva,~N.~D.; Windus,~T.~L.; Gordon,~M.~S.; Moyer,~B.~A.; Bryantsev,~V.~S.; Hay,~B.~P. Computer-aided molecular design of bis-phosphine oxide lanthanide extractants. \emph{Inorganic Chemistry} \textbf{2016}, \emph{55}, 5787--5803\relax
\mciteBstWouldAddEndPuncttrue
\mciteSetBstMidEndSepPunct{\mcitedefaultmidpunct}
{\mcitedefaultendpunct}{\mcitedefaultseppunct}\relax
\EndOfBibitem
\bibitem[Ellis \latin{et~al.}(2017)Ellis, Brigham, Delmau, Ivanov, Williams, Vo, Reinhart, Moyer, and Bryantsev]{ellis2017straining}
Ellis,~R.~J.; Brigham,~D.~M.; Delmau,~L.; Ivanov,~A.~S.; Williams,~N.~J.; Vo,~M.~N.; Reinhart,~B.; Moyer,~B.~A.; Bryantsev,~V.~S. “Straining” to separate the rare earths: how the lanthanide contraction impacts chelation by diglycolamide ligands. \emph{Inorganic Chemistry} \textbf{2017}, \emph{56}, 1152--1160\relax
\mciteBstWouldAddEndPuncttrue
\mciteSetBstMidEndSepPunct{\mcitedefaultmidpunct}
{\mcitedefaultendpunct}{\mcitedefaultseppunct}\relax
\EndOfBibitem
\bibitem[Zhang \latin{et~al.}(2022)Zhang, Adelman, Arko, De~Silva, Su, Kozimor, Mocko, Shafer, Stein, Schreckenbach, and et~al.]{zhang2022advancing}
Zhang,~X.; Adelman,~S.~L.; Arko,~B.~T.; De~Silva,~C.~R.; Su,~J.; Kozimor,~S.~A.; Mocko,~V.; Shafer,~J.~C.; Stein,~B.~W.; Schreckenbach,~G.; et~al. Advancing the Am Extractant Design through the Interplay among Planarity, Preorganization, and Substitution Effects. \emph{Inorganic Chemistry} \textbf{2022}, \emph{61}, 11556--11570\relax
\mciteBstWouldAddEndPuncttrue
\mciteSetBstMidEndSepPunct{\mcitedefaultmidpunct}
{\mcitedefaultendpunct}{\mcitedefaultseppunct}\relax
\EndOfBibitem
\bibitem[Laio and Parrinello(2002)Laio, and Parrinello]{laio2002escaping}
Laio,~A.; Parrinello,~M. Escaping free-energy minima. \emph{Proceedings of the national academy of sciences} \textbf{2002}, \emph{99}, 12562--12566\relax
\mciteBstWouldAddEndPuncttrue
\mciteSetBstMidEndSepPunct{\mcitedefaultmidpunct}
{\mcitedefaultendpunct}{\mcitedefaultseppunct}\relax
\EndOfBibitem
\bibitem[Barducci \latin{et~al.}(2011)Barducci, Bonomi, and Parrinello]{barducci2011metadynamics}
Barducci,~A.; Bonomi,~M.; Parrinello,~M. Metadynamics. \emph{Wiley Interdisciplinary Reviews: Computational Molecular Science} \textbf{2011}, \emph{1}, 826--843\relax
\mciteBstWouldAddEndPuncttrue
\mciteSetBstMidEndSepPunct{\mcitedefaultmidpunct}
{\mcitedefaultendpunct}{\mcitedefaultseppunct}\relax
\EndOfBibitem
\bibitem[Valsson \latin{et~al.}(2016)Valsson, Tiwary, and Parrinello]{valsson2016enhancing}
Valsson,~O.; Tiwary,~P.; Parrinello,~M. Enhancing important fluctuations: Rare events and metadynamics from a conceptual viewpoint. \emph{Annual review of physical chemistry} \textbf{2016}, \emph{67}, 159--184\relax
\mciteBstWouldAddEndPuncttrue
\mciteSetBstMidEndSepPunct{\mcitedefaultmidpunct}
{\mcitedefaultendpunct}{\mcitedefaultseppunct}\relax
\EndOfBibitem
\bibitem[Lumetta \latin{et~al.}(2002)Lumetta, Rapko, Garza, Hay, Gilbertson, Weakley, and Hutchison]{lumetta2002deliberate}
Lumetta,~G.~J.; Rapko,~B.~M.; Garza,~P.~A.; Hay,~B.~P.; Gilbertson,~R.~D.; Weakley,~T.~J.; Hutchison,~J.~E. Deliberate design of ligand architecture yields dramatic enhancement of metal ion affinity. \emph{Journal of the American Chemical Society} \textbf{2002}, \emph{124}, 5644--5645\relax
\mciteBstWouldAddEndPuncttrue
\mciteSetBstMidEndSepPunct{\mcitedefaultmidpunct}
{\mcitedefaultendpunct}{\mcitedefaultseppunct}\relax
\EndOfBibitem
\bibitem[Merrill \latin{et~al.}(2011)Merrill, Harrington, Lee, and Hancock]{merrill2011unusual}
Merrill,~D.; Harrington,~J.~M.; Lee,~H.-S.; Hancock,~R.~D. Unusual metal ion selectivities of the highly preorganized tetradentrate ligand 1, 10-phenanthroline-2, 9-dicarboxamide: a thermodynamic and fluorescence study. \emph{Inorganic chemistry} \textbf{2011}, \emph{50}, 8348--8355\relax
\mciteBstWouldAddEndPuncttrue
\mciteSetBstMidEndSepPunct{\mcitedefaultmidpunct}
{\mcitedefaultendpunct}{\mcitedefaultseppunct}\relax
\EndOfBibitem
\bibitem[Merrill and Hancock(2011)Merrill, and Hancock]{merrill2011metal}
Merrill,~D.; Hancock,~R.~D. Metal ion selectivities of the highly preorganized tetradentate ligand 1, 10-phenanthroline-2, 9-dicarboxamide with lanthanide (III) ions and some actinide ions. \emph{Radiochimica Acta} \textbf{2011}, \emph{99}, 161--166\relax
\mciteBstWouldAddEndPuncttrue
\mciteSetBstMidEndSepPunct{\mcitedefaultmidpunct}
{\mcitedefaultendpunct}{\mcitedefaultseppunct}\relax
\EndOfBibitem
\bibitem[Gephart~III \latin{et~al.}(2009)Gephart~III, Williams, Reibenspies, De~Sousa, and Hancock]{gephart2009complexation}
Gephart~III,~R.~T.; Williams,~N.~J.; Reibenspies,~J.~H.; De~Sousa,~A.~S.; Hancock,~R.~D. Complexation of metal ions of higher charge by the highly preorganized tetradentate ligand 2, 9-bis (hydroxymethyl)-1, 10-phenanthroline. A crystallographic and thermodynamic study. \emph{Inorganic chemistry} \textbf{2009}, \emph{48}, 8201--8209\relax
\mciteBstWouldAddEndPuncttrue
\mciteSetBstMidEndSepPunct{\mcitedefaultmidpunct}
{\mcitedefaultendpunct}{\mcitedefaultseppunct}\relax
\EndOfBibitem
\bibitem[Williams \latin{et~al.}(2009)Williams, Dean, VanDerveer, Luckay, and Hancock]{williams2009strong}
Williams,~N.~J.; Dean,~N.~E.; VanDerveer,~D.~G.; Luckay,~R.~C.; Hancock,~R.~D. Strong metal ion size based selectivity of the highly preorganized ligand PDA (1, 10-phenanthroline-2, 9-dicarboxylic acid) with trivalent metal ions. A crystallographic, fluorometric, and thermodynamic Study. \emph{Inorganic chemistry} \textbf{2009}, \emph{48}, 7853--7863\relax
\mciteBstWouldAddEndPuncttrue
\mciteSetBstMidEndSepPunct{\mcitedefaultmidpunct}
{\mcitedefaultendpunct}{\mcitedefaultseppunct}\relax
\EndOfBibitem
\bibitem[Lashley \latin{et~al.}(2016)Lashley, Ivanov, Bryantsev, Dai, and Hancock]{lashley2016highly}
Lashley,~M.~A.; Ivanov,~A.~S.; Bryantsev,~V.~S.; Dai,~S.; Hancock,~R.~D. Highly preorganized ligand 1, 10-phenanthroline-2, 9-dicarboxylic acid for the selective recovery of uranium from seawater in the presence of competing vanadium species. \emph{Inorganic Chemistry} \textbf{2016}, \emph{55}, 10818--10829\relax
\mciteBstWouldAddEndPuncttrue
\mciteSetBstMidEndSepPunct{\mcitedefaultmidpunct}
{\mcitedefaultendpunct}{\mcitedefaultseppunct}\relax
\EndOfBibitem
\bibitem[McNamara \latin{et~al.}(1999)McNamara, Lumetta, and Rapko]{mcnamara1999extraction}
McNamara,~B.~K.; Lumetta,~G.~J.; Rapko,~B.~M. Extraction of europium (III) ion with tetrahexylmalonamides. \emph{Solvent Extraction and Ion Exchange} \textbf{1999}, \emph{17}, 1403--1421\relax
\mciteBstWouldAddEndPuncttrue
\mciteSetBstMidEndSepPunct{\mcitedefaultmidpunct}
{\mcitedefaultendpunct}{\mcitedefaultseppunct}\relax
\EndOfBibitem
\bibitem[Lumetta \latin{et~al.}(1999)Lumetta, McNamara, Rapko, and Hutchison]{lumetta1999complexation}
Lumetta,~G.~J.; McNamara,~B.~K.; Rapko,~B.~M.; Hutchison,~J.~E. Complexation of uranyl ion by tetrahexylmalonamides: an equilibrium modeling and infrared spectroscopic study. \emph{Inorganica Chimica Acta} \textbf{1999}, \emph{293}, 195--205\relax
\mciteBstWouldAddEndPuncttrue
\mciteSetBstMidEndSepPunct{\mcitedefaultmidpunct}
{\mcitedefaultendpunct}{\mcitedefaultseppunct}\relax
\EndOfBibitem
\bibitem[Nigond \latin{et~al.}(1995)Nigond, Condamines, Cordier, Livet, Madic, Cuillerdier, Musikas, and Hudson]{nigond1995recent}
Nigond,~L.; Condamines,~N.; Cordier,~P.; Livet,~J.; Madic,~C.; Cuillerdier,~C.; Musikas,~C.; Hudson,~M. Recent advances in the treatment of nuclear wastes by the use of diamide and picolinamide extractants. \emph{Separation Science and Technology} \textbf{1995}, \emph{30}, 2075--2099\relax
\mciteBstWouldAddEndPuncttrue
\mciteSetBstMidEndSepPunct{\mcitedefaultmidpunct}
{\mcitedefaultendpunct}{\mcitedefaultseppunct}\relax
\EndOfBibitem
\bibitem[Philip~Horwitz \latin{et~al.}(1985)Philip~Horwitz, Kalina, Diamond, Vandegrift, and Schulz]{philip1985truex}
Philip~Horwitz,~E.; Kalina,~D.~C.; Diamond,~H.; Vandegrift,~G.~F.; Schulz,~W.~W. The truex process-a process for the extraction of the tkansuranic elements erom nitric AC in wastes utilizing modified purex solvent. \emph{Solvent Extraction and Ion Exchange} \textbf{1985}, \emph{3}, 75--109\relax
\mciteBstWouldAddEndPuncttrue
\mciteSetBstMidEndSepPunct{\mcitedefaultmidpunct}
{\mcitedefaultendpunct}{\mcitedefaultseppunct}\relax
\EndOfBibitem
\bibitem[Zhao and Truhlar(2008)Zhao, and Truhlar]{zhao2008m06}
Zhao,~Y.; Truhlar,~D.~G. The M06 suite of density functionals for main group thermochemistry, thermochemical kinetics, noncovalent interactions, excited states, and transition elements: two new functionals and systematic testing of four M06-class functionals and 12 other functionals. \emph{Theoretical Chemistry Accounts} \textbf{2008}, \emph{120}, 215--241\relax
\mciteBstWouldAddEndPuncttrue
\mciteSetBstMidEndSepPunct{\mcitedefaultmidpunct}
{\mcitedefaultendpunct}{\mcitedefaultseppunct}\relax
\EndOfBibitem
\bibitem[Pritchard \latin{et~al.}(2019)Pritchard, Altarawy, Didier, Gibson, and Windus]{pritchard2019new}
Pritchard,~B.~P.; Altarawy,~D.; Didier,~B.; Gibson,~T.~D.; Windus,~T.~L. New basis set exchange: An open, up-to-date resource for the molecular sciences community. \emph{Journal of Chemical Information and Modeling} \textbf{2019}, \emph{59}, 4814--4820\relax
\mciteBstWouldAddEndPuncttrue
\mciteSetBstMidEndSepPunct{\mcitedefaultmidpunct}
{\mcitedefaultendpunct}{\mcitedefaultseppunct}\relax
\EndOfBibitem
\bibitem[Frisch \latin{et~al.}(2016)Frisch, Trucks, Schlegel, Scuseria, Robb, Cheeseman, Scalmani, Barone, Petersson, Nakatsuji, and et~al.]{g16}
Frisch,~M.~J.; Trucks,~G.~W.; Schlegel,~H.~B.; Scuseria,~G.~E.; Robb,~M.~A.; Cheeseman,~J.~R.; Scalmani,~G.; Barone,~V.; Petersson,~G.~A.; Nakatsuji,~H.; et~al. Gaussian 16 {R}evision {C}.01. 2016; Gaussian Inc. Wallingford CT\relax
\mciteBstWouldAddEndPuncttrue
\mciteSetBstMidEndSepPunct{\mcitedefaultmidpunct}
{\mcitedefaultendpunct}{\mcitedefaultseppunct}\relax
\EndOfBibitem
\bibitem[He \latin{et~al.}(2020)He, Man, Yang, Lee, and Wang]{he2020fast}
He,~X.; Man,~V.~H.; Yang,~W.; Lee,~T.-S.; Wang,~J. A fast and high-quality charge model for the next generation general AMBER force field. \emph{The Journal of Chemical Physics} \textbf{2020}, \emph{153}, 114502\relax
\mciteBstWouldAddEndPuncttrue
\mciteSetBstMidEndSepPunct{\mcitedefaultmidpunct}
{\mcitedefaultendpunct}{\mcitedefaultseppunct}\relax
\EndOfBibitem
\bibitem[Jakalian \latin{et~al.}(2000)Jakalian, Bush, Jack, and Bayly]{jakalian2000fast}
Jakalian,~A.; Bush,~B.~L.; Jack,~D.~B.; Bayly,~C.~I. Fast, efficient generation of high-quality atomic charges. AM1-BCC model: I. Method. \emph{Journal of computational chemistry} \textbf{2000}, \emph{21}, 132--146\relax
\mciteBstWouldAddEndPuncttrue
\mciteSetBstMidEndSepPunct{\mcitedefaultmidpunct}
{\mcitedefaultendpunct}{\mcitedefaultseppunct}\relax
\EndOfBibitem
\bibitem[Jakalian \latin{et~al.}(2002)Jakalian, Jack, and Bayly]{jakalian2002fast}
Jakalian,~A.; Jack,~D.~B.; Bayly,~C.~I. Fast, efficient generation of high-quality atomic charges. AM1-BCC model: II. Parameterization and validation. \emph{Journal of computational chemistry} \textbf{2002}, \emph{23}, 1623--1641\relax
\mciteBstWouldAddEndPuncttrue
\mciteSetBstMidEndSepPunct{\mcitedefaultmidpunct}
{\mcitedefaultendpunct}{\mcitedefaultseppunct}\relax
\EndOfBibitem
\bibitem[Jorgensen \latin{et~al.}(1996)Jorgensen, Maxwell, and Tirado-Rives]{jorgensen1996development}
Jorgensen,~W.~L.; Maxwell,~D.~S.; Tirado-Rives,~J. Development and testing of the OPLS all-atom force field on conformational energetics and properties of organic liquids. \emph{Journal of the American Chemical Society} \textbf{1996}, \emph{118}, 11225--11236\relax
\mciteBstWouldAddEndPuncttrue
\mciteSetBstMidEndSepPunct{\mcitedefaultmidpunct}
{\mcitedefaultendpunct}{\mcitedefaultseppunct}\relax
\EndOfBibitem
\bibitem[Vo \latin{et~al.}(2015)Vo, Hawkins, Dang, Nilsson, and Nguyen]{vo2015computational}
Vo,~Q.~N.; Hawkins,~C.~A.; Dang,~L.~X.; Nilsson,~M.; Nguyen,~H.~D. Computational study of molecular structure and self-association of tri-n-butyl phosphates in n-dodecane. \emph{The Journal of Physical Chemistry B} \textbf{2015}, \emph{119}, 1588--1597\relax
\mciteBstWouldAddEndPuncttrue
\mciteSetBstMidEndSepPunct{\mcitedefaultmidpunct}
{\mcitedefaultendpunct}{\mcitedefaultseppunct}\relax
\EndOfBibitem
\bibitem[Mart{\'\i}nez \latin{et~al.}(2009)Mart{\'\i}nez, Andrade, Birgin, and Mart{\'\i}nez]{martinez2009packmol}
Mart{\'\i}nez,~L.; Andrade,~R.; Birgin,~E.~G.; Mart{\'\i}nez,~J.~M. PACKMOL: A package for building initial configurations for molecular dynamics simulations. \emph{Journal of Computational Chemistry} \textbf{2009}, \emph{30}, 2157--2164\relax
\mciteBstWouldAddEndPuncttrue
\mciteSetBstMidEndSepPunct{\mcitedefaultmidpunct}
{\mcitedefaultendpunct}{\mcitedefaultseppunct}\relax
\EndOfBibitem
\bibitem[Martyna \latin{et~al.}(1994)Martyna, Tobias, and Klein]{martyna1994constant}
Martyna,~G.~J.; Tobias,~D.~J.; Klein,~M.~L. Constant pressure molecular dynamics algorithms. \emph{The Journal of Chemical Physics} \textbf{1994}, \emph{101}, 4177--4189\relax
\mciteBstWouldAddEndPuncttrue
\mciteSetBstMidEndSepPunct{\mcitedefaultmidpunct}
{\mcitedefaultendpunct}{\mcitedefaultseppunct}\relax
\EndOfBibitem
\bibitem[Parrinello and Rahman(1981)Parrinello, and Rahman]{parrinello1981polymorphic}
Parrinello,~M.; Rahman,~A. Polymorphic transitions in single crystals: A new molecular dynamics method. \emph{Journal of Applied physics} \textbf{1981}, \emph{52}, 7182--7190\relax
\mciteBstWouldAddEndPuncttrue
\mciteSetBstMidEndSepPunct{\mcitedefaultmidpunct}
{\mcitedefaultendpunct}{\mcitedefaultseppunct}\relax
\EndOfBibitem
\bibitem[Shinoda \latin{et~al.}(2004)Shinoda, Shiga, and Mikami]{shinoda2004rapid}
Shinoda,~W.; Shiga,~M.; Mikami,~M. Rapid estimation of elastic constants by molecular dynamics simulation under constant stress. \emph{Physical Review B} \textbf{2004}, \emph{69}, 134103\relax
\mciteBstWouldAddEndPuncttrue
\mciteSetBstMidEndSepPunct{\mcitedefaultmidpunct}
{\mcitedefaultendpunct}{\mcitedefaultseppunct}\relax
\EndOfBibitem
\bibitem[Dasari and Mallik(2020)Dasari, and Mallik]{dasari2020conformational}
Dasari,~S.; Mallik,~B.~S. Conformational free-energy landscapes of alanine dipeptide in hydrated ionic liquids from enhanced sampling methods. \emph{The Journal of Physical Chemistry B} \textbf{2020}, \emph{124}, 6728--6737\relax
\mciteBstWouldAddEndPuncttrue
\mciteSetBstMidEndSepPunct{\mcitedefaultmidpunct}
{\mcitedefaultendpunct}{\mcitedefaultseppunct}\relax
\EndOfBibitem
\bibitem[Thompson \latin{et~al.}(2022)Thompson, Aktulga, Berger, Bolintineanu, Brown, Crozier, in't Veld, Kohlmeyer, Moore, Nguyen, and et~al]{thompson2022lammps}
Thompson,~A.~P.; Aktulga,~H.~M.; Berger,~R.; Bolintineanu,~D.~S.; Brown,~W.~M.; Crozier,~P.~S.; in't Veld,~P.~J.; Kohlmeyer,~A.; Moore,~S.~G.; Nguyen,~T.~D.; et~al LAMMPS-a flexible simulation tool for particle-based materials modeling at the atomic, meso, and continuum scales. \emph{Computer Physics Communications} \textbf{2022}, \emph{271}, 108171\relax
\mciteBstWouldAddEndPuncttrue
\mciteSetBstMidEndSepPunct{\mcitedefaultmidpunct}
{\mcitedefaultendpunct}{\mcitedefaultseppunct}\relax
\EndOfBibitem
\bibitem[Tribello \latin{et~al.}(2014)Tribello, Bonomi, Branduardi, Camilloni, and Bussi]{tribello2014plumed}
Tribello,~G.~A.; Bonomi,~M.; Branduardi,~D.; Camilloni,~C.; Bussi,~G. PLUMED 2: New feathers for an old bird. \emph{Computer Physics Communications} \textbf{2014}, \emph{185}, 604--613\relax
\mciteBstWouldAddEndPuncttrue
\mciteSetBstMidEndSepPunct{\mcitedefaultmidpunct}
{\mcitedefaultendpunct}{\mcitedefaultseppunct}\relax
\EndOfBibitem
\bibitem[Ramachandran \latin{et~al.}(1963)Ramachandran, Ramakrishnan, and Sasisekharan]{RAMACHANDRAN196395}
Ramachandran,~G.; Ramakrishnan,~C.; Sasisekharan,~V. Stereochemistry of polypeptide chain configurations. \emph{Journal of Molecular Biology} \textbf{1963}, \emph{7}, 95--99\relax
\mciteBstWouldAddEndPuncttrue
\mciteSetBstMidEndSepPunct{\mcitedefaultmidpunct}
{\mcitedefaultendpunct}{\mcitedefaultseppunct}\relax
\EndOfBibitem
\bibitem[Ellis and Antonio(2012)Ellis, and Antonio]{ellis2012coordination}
Ellis,~R.~J.; Antonio,~M.~R. Coordination structures and supramolecular architectures in a cerium (III)--malonamide solvent extraction system. \emph{Langmuir} \textbf{2012}, \emph{28}, 5987--5998\relax
\mciteBstWouldAddEndPuncttrue
\mciteSetBstMidEndSepPunct{\mcitedefaultmidpunct}
{\mcitedefaultendpunct}{\mcitedefaultseppunct}\relax
\EndOfBibitem
\bibitem[Ellis \latin{et~al.}(2013)Ellis, Meridiano, Chiarizia, Berthon, Muller, Couston, and Antonio]{ellis2013periodic}
Ellis,~R.~J.; Meridiano,~Y.; Chiarizia,~R.; Berthon,~L.; Muller,~J.; Couston,~L.; Antonio,~M.~R. Periodic behavior of lanthanide coordination within reverse micelles. \emph{Chemistry--A European Journal} \textbf{2013}, \emph{19}, 2663--2675\relax
\mciteBstWouldAddEndPuncttrue
\mciteSetBstMidEndSepPunct{\mcitedefaultmidpunct}
{\mcitedefaultendpunct}{\mcitedefaultseppunct}\relax
\EndOfBibitem
\bibitem[Ellis \latin{et~al.}(2014)Ellis, Meridiano, Muller, Berthon, Guilbaud, Zorz, Antonio, Demars, and Zemb]{ellis2014complexation}
Ellis,~R.~J.; Meridiano,~Y.; Muller,~J.; Berthon,~L.; Guilbaud,~P.; Zorz,~N.; Antonio,~M.~R.; Demars,~T.; Zemb,~T. Complexation-Induced Supramolecular Assembly Drives Metal-Ion Extraction. \emph{Chemistry--A European Journal} \textbf{2014}, \emph{20}, 12796--12807\relax
\mciteBstWouldAddEndPuncttrue
\mciteSetBstMidEndSepPunct{\mcitedefaultmidpunct}
{\mcitedefaultendpunct}{\mcitedefaultseppunct}\relax
\EndOfBibitem
\bibitem[Kravchuk \latin{et~al.}(2023)Kravchuk, Wang, Servis, and Wilson]{kravchuk2023structural}
Kravchuk,~D.; Wang,~X.; Servis,~M.; Wilson,~R. Structural Trends and Vibrational Analysis of N, N, N', N'-Tetramethylmalonamide Complexes Across the Lanthanide Series. \emph{ChemRxiv [Preprint]} \textbf{2023}, \relax
\mciteBstWouldAddEndPunctfalse
\mciteSetBstMidEndSepPunct{\mcitedefaultmidpunct}
{}{\mcitedefaultseppunct}\relax
\EndOfBibitem
\bibitem[Berthon \latin{et~al.}(2007)Berthon, Martinet, Testard, Madic, and Zemb]{berthon2007solvent}
Berthon,~L.; Martinet,~L.; Testard,~F.; Madic,~C.; Zemb,~T. Solvent penetration and sterical stabilization of reverse aggregates based on the DIAMEX process extracting molecules: Consequences for the third phase formation. \emph{Solvent Extraction and Ion Exchange} \textbf{2007}, \emph{25}, 545--576\relax
\mciteBstWouldAddEndPuncttrue
\mciteSetBstMidEndSepPunct{\mcitedefaultmidpunct}
{\mcitedefaultendpunct}{\mcitedefaultseppunct}\relax
\EndOfBibitem
\bibitem[Bonnett \latin{et~al.}(2023)Bonnett, Sheyfer, Wimalasiri, Nayak, Lal, Zhang, Seifert, Stephenson, and Servis]{bonnett2023critical}
Bonnett,~B.~L.; Sheyfer,~D.; Wimalasiri,~P.~N.; Nayak,~S.; Lal,~J.; Zhang,~Q.; Seifert,~S.; Stephenson,~G.~B.; Servis,~M.~J. Critical fluctuations in liquid--liquid extraction organic phases controlled by extractant and diluent molecular structure. \emph{Physical Chemistry Chemical Physics} \textbf{2023}, \emph{25}, 16389--16403\relax
\mciteBstWouldAddEndPuncttrue
\mciteSetBstMidEndSepPunct{\mcitedefaultmidpunct}
{\mcitedefaultendpunct}{\mcitedefaultseppunct}\relax
\EndOfBibitem
\bibitem[Erlinger \latin{et~al.}(1998)Erlinger, Gazeau, Zemb, Madic, Lefrancois, Hebrant, and Tondre]{erlinger1998effect}
Erlinger,~C.; Gazeau,~D.; Zemb,~T.; Madic,~C.; Lefrancois,~L.; Hebrant,~M.; Tondre,~C. Effect of nitric acid extraction on phase behavior, microstructure and interactions between primary aggregates in the system dimethyldibutyltetradecylmalonamide (DMDBTDMA)/n-dodecane/water: A phase analysis and small angle X-ray scattering (SAXS) characterisation study. \emph{Solvent Extraction and Ion Exchange} \textbf{1998}, \emph{16}, 707--738\relax
\mciteBstWouldAddEndPuncttrue
\mciteSetBstMidEndSepPunct{\mcitedefaultmidpunct}
{\mcitedefaultendpunct}{\mcitedefaultseppunct}\relax
\EndOfBibitem
\bibitem[Makombe \latin{et~al.}(2022)Makombe, Bourgeois, Berthon, and Meyer]{makombe2022uranium}
Makombe,~E.; Bourgeois,~D.; Berthon,~L.; Meyer,~D. Uranium (VI) and thorium (IV) extraction by malonamides: Impact of ligand molecular topology on selectivity. \emph{Journal of Molecular Liquids} \textbf{2022}, \emph{368}, 120701\relax
\mciteBstWouldAddEndPuncttrue
\mciteSetBstMidEndSepPunct{\mcitedefaultmidpunct}
{\mcitedefaultendpunct}{\mcitedefaultseppunct}\relax
\EndOfBibitem
\bibitem[Johnson \latin{et~al.}(2023)Johnson, Driscoll, Damron, Ivanov, and Jansone-Popova]{johnson2023size}
Johnson,~K.~R.; Driscoll,~D.~M.; Damron,~J.~T.; Ivanov,~A.~S.; Jansone-Popova,~S. Size Selective Ligand Tug of War Strategy to Separate Rare Earth Elements. \emph{JACS Au} \textbf{2023}, \emph{3}, 584--591\relax
\mciteBstWouldAddEndPuncttrue
\mciteSetBstMidEndSepPunct{\mcitedefaultmidpunct}
{\mcitedefaultendpunct}{\mcitedefaultseppunct}\relax
\EndOfBibitem
\bibitem[Kalina \latin{et~al.}(1981)Kalina, Horwitz, Kaplan, and Muscatello]{kalina1981extraction}
Kalina,~D.; Horwitz,~E.; Kaplan,~L.; Muscatello,~A. The extraction of Am (III) and Fe (III) by selected dihexyl N, N-dialkylcarbamoylmethyl-phosphonates,-phosphinates and-phosphine oxides from nitrate media. \emph{Separation Science and Technology} \textbf{1981}, \emph{16}, 1127--1145\relax
\mciteBstWouldAddEndPuncttrue
\mciteSetBstMidEndSepPunct{\mcitedefaultmidpunct}
{\mcitedefaultendpunct}{\mcitedefaultseppunct}\relax
\EndOfBibitem
\bibitem[Horwitz \latin{et~al.}(1982)Horwitz, Kalina, Kaplan, Mason, and Diamond]{horwitz1982selected}
Horwitz,~E.; Kalina,~D.; Kaplan,~L.; Mason,~G.; Diamond,~H. Selected alkyl (phenyl)-N, N-dialkylcarbamoylmethylphosphine oxides as extractants for Am (III) from nitric acid media. \emph{Separation Science and Technology} \textbf{1982}, \emph{17}, 1261--1279\relax
\mciteBstWouldAddEndPuncttrue
\mciteSetBstMidEndSepPunct{\mcitedefaultmidpunct}
{\mcitedefaultendpunct}{\mcitedefaultseppunct}\relax
\EndOfBibitem
\bibitem[Ensing \latin{et~al.}(2005)Ensing, Laio, Parrinello, and Klein]{ensing2005recipe}
Ensing,~B.; Laio,~A.; Parrinello,~M.; Klein,~M.~L. A recipe for the computation of the free energy barrier and the lowest free energy path of concerted reactions. \emph{The Journal of Physical Chemistry B} \textbf{2005}, \emph{109}, 6676--6687\relax
\mciteBstWouldAddEndPuncttrue
\mciteSetBstMidEndSepPunct{\mcitedefaultmidpunct}
{\mcitedefaultendpunct}{\mcitedefaultseppunct}\relax
\EndOfBibitem
\bibitem[Fu \latin{et~al.}(2020)Fu, Chen, Wang, Chai, Shao, Cai, and Chipot]{fu2020finding}
Fu,~H.; Chen,~H.; Wang,~X.; Chai,~H.; Shao,~X.; Cai,~W.; Chipot,~C. Finding an optimal pathway on a multidimensional free-energy landscape. \emph{Journal of Chemical Information and Modeling} \textbf{2020}, \emph{60}, 5366--5374\relax
\mciteBstWouldAddEndPuncttrue
\mciteSetBstMidEndSepPunct{\mcitedefaultmidpunct}
{\mcitedefaultendpunct}{\mcitedefaultseppunct}\relax
\EndOfBibitem
\bibitem[Henkelman and J{\'o}nsson(2000)Henkelman, and J{\'o}nsson]{henkelman2000improved}
Henkelman,~G.; J{\'o}nsson,~H. Improved tangent estimate in the nudged elastic band method for finding minimum energy paths and saddle points. \emph{The Journal of chemical physics} \textbf{2000}, \emph{113}, 9978--9985\relax
\mciteBstWouldAddEndPuncttrue
\mciteSetBstMidEndSepPunct{\mcitedefaultmidpunct}
{\mcitedefaultendpunct}{\mcitedefaultseppunct}\relax
\EndOfBibitem
\end{mcitethebibliography}

\end{document}